\documentclass[]{article}

\usepackage{color}
\usepackage{amsmath}
\usepackage{mathtools}
\usepackage{fullpage}
\usepackage[]{algorithm2e}
\usepackage{algorithmic}
\DeclareMathOperator*{\argmin}{argmin}
\algsetup{linenosize=\small}
\usepackage[table,xcdraw]{xcolor}
\usepackage{multirow}
\usepackage[super]{nth}
\usepackage{graphicx}
\usepackage{caption}
\usepackage{subcaption}
\usepackage{setspace}
\usepackage{textcomp}
\usepackage{xspace}
\usepackage{siunitx}
\usepackage{epsfig}
\usepackage{epstopdf}
\usepackage{soul}
\usepackage{url}
\usepackage{tablefootnote}
\usepackage[labelformat=simple]{subcaption}

\let\OldTexttrademark\texttrademark
\renewcommand{\texttrademark}{\OldTexttrademark\xspace}%

\DeclarePairedDelimiter\ceil{\lceil}{\rceil}
\usepackage{graphicx}
\usepackage{amssymb}
\usepackage{authblk}
\providecommand{\keywords}[1]{\textbf{\textit{Keywords:}} #1}

\title{Multi-hop Communication in the Uplink for LPWANs}
\author{Sergio Barrachina-Mu\~noz, Boris Bellalta, Toni Adame, and Albert Bel}
\affil{Dept. of Information and Communication Technologies \\ Universitat Pompeu Fabra (UPF), Barcelona}
\date{}

\begin{document}
	
	\maketitle
	
	\begin{abstract}
		%% Text of abstract
		Low-Power Wide Area Networks (LPWANs) have arisen as a promising communication technology for supporting Internet of Things (IoT) services due to their low power operation, wide coverage range, low cost and scalability. However, most LPWAN solutions like SIGFOX\texttrademark or LoRaWAN\texttrademark rely on star topology networks, where stations (STAs) transmit directly to the gateway (GW), which often leads to rapid battery depletion in STAs located far from it.
		
		In this work, we analyze the impact on LPWANs energy consumption of multi-hop communication in the uplink, allowing STAs to transmit data packets in lower power levels and higher data rates to closer parent STAs, reducing their energy consumption consequently. To that aim, we introduce the Distance-Ring Exponential Stations Generator (DRESG) framework\footnote{All of the source code in DRESG framework for LPWANs is open \cite{barrachina2017dresg}, encouraging sharing of algorithms between contributors and providing the ability for people to improve on the work of others under the GNU General Public License v3.0.}, designed to evaluate the performance of the so-called optimal-hop routing model, which establishes optimal routing connections in terms of energy efficiency, aiming to balance the consumption among all the STAs in the network. Results show that enabling such multi-hop connections entails higher network lifetimes, reducing significantly the bottleneck consumption in LPWANs with up to thousands of STAs. These results lead to foresee multi-hop communication in the uplink as a promising routing alternative for extending the lifetime of LPWAN deployments.	
	\end{abstract}
	
	
	\section{Introduction} \label{introduction}
	
	% What is IoT, relevance
	The Internet of Things (IoT) concept refers to the interconnection of \textit{things}, i.e., everyday physical objects or entities uniquely identified, embeddable, and provided with the capability of transferring data. In the coming years lots of miscellaneous \textit{things} such as fridges, cars, traffic lights, or even persons will be interconnected. Specifically, it is forecast that there will be approximately 28 billion connected devices by 2021 \cite{ericsson2016}. Assisted driving, environmental monitoring, health-care, or smart environments (e.g., homes, offices, museums, industries, etc.) are just some of the huge number of applications that IoT will make possible \cite{Atzori20102787}.
	
	% How were IoT services supported? 1) Multi-hop short range networks, 2) Cellular networks
	In recent years, two types of communication technologies have been used in IoT deployments: \textit{i}) multi-hop short range networks and, \textit{ii}) traditional cellular networks. On the one hand, multi-hop short-range transmission technologies, such as IEEE 802.15.4-based low-rate wireless personal area networks (e.g., Zigbee), or Bluetooth (formerly standardized as IEEE 802.15.1), provide very low power consumption, which is a critical requirement as \textit{things} are usually battery-driven devices expected to remain operative up to years. However, these technologies were designed for short-range  scenarios (about 100 meters), such as rooms or small buildings, having important limitations in typical IoT deployments requiring wide coverage areas. On the other hand, cellular networks are able to provide ubiquitousness and extensive coverage range (from 1 to 15 km in urban areas). Nonetheless, even though future mobile networks are intended to support machine-to-machine (M2M) services, their operational cost, high power consumption and lack of scalability for massive amount of devices remain major constraints \cite{biral2015challenges}.
	
	% LPWAN arise as a solution to IoT
	Low-Power Wide Area Networks (LPWANs) have arisen as a promising complementary communication technology for IoT. LPWANs are wireless wide area networks designed for achieving large coverage ranges, extending end devices battery lifetime and reducing the operational cost of traditional cellular networks. They are characterized by exploiting the sub-1GHz unlicensed, industrial, scientific and medical (ISM) frequency band, and by sporadically transmitting small packets at low data rates, which leads to achieving very low receptor sensitivities. Therefore, LPWANs are expected to be completely suitable for supporting IoT services, which commonly require low data throughput communications and large coverage ranges.
	
	% Contribution: Star topology in LPWANs and multi-hop possibility
	Most LPWANs like SIGFOX\texttrademark \cite{sigfox2016main} or LoRaWAN\texttrademark \cite{alliance2016lora} are built following a star topology, where end devices, or stations (STAs), are connected directly to the base station (BS) or gateway (GW). This single-hop approach simplifies the network design and provides a robust centralized control of the network while allowing the use of plain protocol stacks (from channel access to routing). However, in such topologies, STAs rely deeply on their transceiver's capabilities (i.e., transmission power, antenna gain, data rate, etc.) as they are intended to reach the GW directly. This strong requirement may lead to rapid energy consumption in STAs located far from the GW as they are required to transmit in high power levels, shortening their lifetime as a consequence. Moreover, single-hop topologies hinder the inclusion of devices with transmission power limitations because of such range constraint.
	
	% Paper contributions
	In this work, we analyze the impact on LPWANs energy consumption of enabling multi-hop routing connections in the uplink. To that aim, we introduce the Distance-Ring Exponential Stations Generator (DRESG) framework \cite{barrachina2017dresg} to evaluate the performance of the so-called optimal-hop routing model, which establishes optimal routing paths in terms of energy saving by means of balancing the consumption among all the STAs in the network. Results show that for LPWANs of up to several thousands of STAs, enabling such multi-hop connections in the uplink leads to higher network lifetimes than with single-hop transmissions since the consumption of STAs located far from the GW is significantly reduced. Results also show that such improvements are deeply enhanced with data aggregation, and that the benefits of multi-hop communication in the uplink are independent of the type of transceiver considered. 
	
	% Instead, for networks with a larger number of STAs, single-hop connections may be the best option in order to avoid consumption bottlenecks at parent STAs.
	
	% Remainder of paper organization
	The remainder of this paper is organized as follows: Section \ref{sec:lpwans} describes the main features, solutions and topologies of LPWANs. Section \ref{sec:considerations} depicts the DRESG framework and the assumed considerations. Routing models in DRESG and the corresponding algorithm for computing the optimal-hop paths are detailed in Section \ref{sec:routing_impact}. The routing model impact on energy in different DRESG scenarios is evaluated in Section \ref{sec:performance_evaluation}. Lastly, Section \ref{sec:conclusions} presents the conclusions and next steps.
	
	%%%%%%%%%%%%%%%%%%%%%%%%%%%%%%%%%%%%%%%%%%%%%%%%
	%%%%%%  II. LOW POWER WIDE AREA NETWORKS  %%%%%%
	%%%%%%%%%%%%%%%%%%%%%%%%%%%%%%%%%%%%%%%%%%%%%%%%
	\section{Low Power Wide Area Networks}	\label{sec:lpwans}
	Several LPWAN solutions have emerged to support IoT and M2M services. In this section we depict their main common features and singularities, focusing on the supported network topologies.
	
	%%% Common LPWAN features
	\subsection{Common features}
	As any communication technology, an LPWAN deployment may be different depending on the selected standard and scenario. However, there are four main common features strongly tied to LPWANs:
	\begin{itemize}
		\item \textbf{Low-power consumption:} LPWANs are generally composed of STAs with limited energy resources as they are usually small and battery-powered devices. Hence, as the lifetime of such STAs is determined by their power consumption and battery capacity, energy-aware designs are a priority; even more when, in most of IoT deployments, manually replacing the batteries when depleted entails huge logistical expenses due to accessibility restrictions. Generally, LPWAN's are low power operation as they support applications requiring sporadic transmission of small data packets. Also, simplified protocol stacks and improved sensitivities lead to lower power consumption than in traditional cellular networks. In \cite{margelis2015low}, authors discuss the suitability of low throughput networks like LoRaWAN\texttrademark and SIGFOX\texttrademark on different applications.
		
		\item \textbf{Wide coverage range:} larger coverage ranges (1 to 15 km in urban areas) than in multi-hop short range networks are achievable due to the use of hardware supporting very low sensitivities (below -140 dBm \cite{rpma}), coming at the price of requiring low data rates of up to hundreds of bits per second.
		
		\item \textbf{Low cost:} Gartner\footnote{Gartner Inc. website: \url{http://www.gartner.com/technology/home.jsp}} forecasts that long-term LPWANs will support many hundreds of thousands of STAs connected to a GW, BS or equivalent \cite{jones2016top}. With such high node density, reducing STAs unit cost is a major priority as the average revenue per unit is expected to be very low.
		
		\item \textbf{Scalability:} when deploying new LPWANs it must be taken into account future user capacity demands (i.e., more STAs per GW) and the potential appearance of other networks placed in near locations. Therefore, the frequency spectrum or even the infrastructure may be shared among networks, which could lead to service downgrades if not properly prevented. In that sense, mechanisms included in the LPWAN protocol stack could cope with such coexistence challenges.
	\end{itemize}
	
	%%% Technologies
	\subsection{Solutions}
	% Different LPWAN solutions

	In Table \ref{table:lpwan_tech} some of the main LPWAN solutions are summarized.
	
	\begin{itemize}
		\item \textbf{SIGFOX\texttrademark}: this technology relies on BS antennas installed on towers of network operators such as Cellnex Telecom\footnote{Cellnex Telecom website: \url{https://www.cellnextelecom.com/}} in Spain or Nettrotter\footnote{Nettrotter website: \url{http://www.nettrotter.io/}} in Italy, providing an end-to-end solution including the actual LPWAN, i.e., STAs plus a certified modem (or GW), and the backend cloud management platform for configuring callbacks to receive and send messages \cite{sigfox2016main}. SIGFOX\texttrademark operates at 868 MHz in the European Union and 902 MHz in the United Sates, dividing the spectrum in 400 channels of 100Hz (from 868.180 to 868.220 MHz in the first case) \cite{margelis2015low}.
		
		Packets in SIGFOX\texttrademark are very small (12 and 8 bytes in uplink and downlink, respectively) and are transmitted at low rate (100 bps) using an ultra-narrow band (UNB) binary phase-shift keying (BPSK). STAs can send up to 140 messages per day, while the GW is allowed to transmit just up to 4 messages.
		
		In SIGFOX\texttrademark no signaling, nor handshakings among STAs and GWs are required, leading to a simple and robust network layer. Regarding uplink communication, STAs willing to transmit a packet select a pseudo-random frequency channel and transmit directly to the GW. For the downlink, STAs should specifically request for network updates, which allows them to be in low power states and save energy most of the time.
		
			% LPWAN SOLUTIONS TABLE: 
		\begin{table}
			\scriptsize
			\centering
			\begin{tabular}{|
					>{\columncolor[HTML]{C0C0C0}}c |c|c|c|c|c|}
				\hline
				\textbf{Solution}                                                    & \cellcolor[HTML]{C0C0C0}\textbf{SIGFOX\texttrademark}                        & \cellcolor[HTML]{C0C0C0}\textbf{LoRaWAN\texttrademark}                                            & \cellcolor[HTML]{C0C0C0}\textbf{Ingenu\texttrademark}                        & \cellcolor[HTML]{C0C0C0}\textbf{D7AP}                                       & \cellcolor[HTML]{C0C0C0}\textbf{HARE\tablefootnote{HARE is a protocol stack and does not require any specific transceiver model for its implementation. In the table, values obtained in tests performed for Zolertia RE-Mote and TelosB by authors in \cite{adame2017hare} are presented.}}                     \\ \hline
				\textbf{Frequency}                                                   & \begin{tabular}[c]{@{}c@{}}868 MHz\\ 902 MHz\end{tabular}      & \begin{tabular}[c]{@{}c@{}}433 MHz\\ 780 MHz\\ 868 MHz\\ 915 MHz\end{tabular}       & 2.4 GHz                                                        & \begin{tabular}[c]{@{}c@{}}433 MHz\\ 868 MHz\\ 915 MHz\end{tabular}         & \begin{tabular}[c]{@{}c@{}}868 MHz\\ 2.4 GHz\end{tabular} \\ \hline
				\textbf{Urban range}                                                 & 10 km                                                          & 15 km                                                                               & 1 to 3 km                                                      & 10 km                                                                       & Up to several km                                                         \\ \hline
				\textbf{\begin{tabular}[c]{@{}c@{}}Uplink \\ data rate\end{tabular}} & 100 bps                                                        & 0.3 to 50 kbps                                                                      & \begin{tabular}[c]{@{}c@{}}624 kbps\\ per sector\end{tabular}  & \begin{tabular}[c]{@{}c@{}}9.6 kbps\\ 55.55 kbps\\ 166.67 kbps\end{tabular} & \begin{tabular}[c]{@{}c@{}}50 kbps\\ 250 kbps\end{tabular}                                                        \\ \hline
				\textbf{Packet size}                                                 & 12 bytes                                                       & user-defined                                                                        & \begin{tabular}[c]{@{}c@{}}6 bytes to\\ 10 kbytes\end{tabular} & $\leq 256$ bytes                                                            & $\leq 127$ bytes                                              \\ \hline
				\textbf{\begin{tabular}[c]{@{}c@{}}STA power\\ output\end{tabular}}  & $\leq$ 20 dBm                                                & \begin{tabular}[c]{@{}c@{}} $\leq$14 dBm (EU)\\  $\leq$27 dBm (US)\end{tabular} &  $\leq$20 dBm                                                & \begin{tabular}[c]{@{}c@{}}FCC/ETSI\\ regulations\end{tabular}              &  $\leq$14 dBm                                                         \\ \hline
				\textbf{Topology}                                                    & Star                                                           & Star-of-stars                                                                       & \begin{tabular}[c]{@{}c@{}}Star\\ Tree*\end{tabular}           & \begin{tabular}[c]{@{}c@{}}Star\\ Tree\end{tabular}                         & \begin{tabular}[c]{@{}c@{}}Star\\ Tree\end{tabular}       \\ \hline
				\textbf{Stage}                                                       & Scale                                                          & Scale                                                                               & Introduction                                                   & Introduction                                                                & Research                                               \\ \hline
			\end{tabular}
			\caption{Overview of some of the major alternatives for LPWAN deployments (based on \cite{raza2016low, adame2017hare}).}
			\label{table:lpwan_tech}
		\end{table}

		\item \textbf{LoRaWAN\texttrademark:}
		% LoRa PHY layer
		LoRa\textsuperscript{\textregistered} is a wireless modulation based on chirp spread spectrum \cite{springer2000spread} providing large communication ranges while maintaining the low power characteristics of the frequency shifting keying (FSK). Hence, LoRa\textsuperscript{\textregistered} is not a complete LPWAN or protocol stack, but a proprietary physical (PHY) layer wireless component.
		
		% LoRaWAN
		LoRaWAN\texttrademark is the rest of LoRa's protocol stack, primarily composed of the MAC and some elements of the network layer, which relay on the LoRa\textsuperscript{\textregistered} modulation scheme. In LoRaWAN\texttrademark, STAs are not associated with an specific GW as uplink packets, transmitted on different frequency channels and data rates (from 300 bps to 50 kbps), are typically received by multiple GWs, which forward them to a cloud-based server \cite{loraWanSpec}. In order to extend the battery lifetime of STAs, a network server sets the data rate and power output of every STA in the LPWAN by means of an adaptive data rate scheme.
		
		There are three classes of STAs in LoRaWAN\texttrademark \cite{goursaud2015dedicated}: A, B and C. In class A, communication is initiated by the STAs in a way that uplink transmissions trigger two short downlink windows. Transmission slots are scheduled by the STAs whenever they are willing to transmit in an ALOHA-based protocol. STAs of class A are the least power consuming. STAs of class B open extra scheduled reception windows through a synchronization beacon, which allows the server identifying when STAs are listening to the channel. Finally, STAs of class C are always listening with continuously open reception windows that are closed only when transmitting, allowing lower latency but higher energy consumption.
		
		% Ingenu
		\item\textbf{Ingenu\texttrademark:} the main particularity of Ingenu\texttrademark is its use of the 2.4 GHz frequency band allowing higher throughput than common LPWANs in the sub-1GHz band \cite{raza2016low}. Due to high frequency bands generally encounter more propagation losses, this proprietary LPWAN solution reaches large ranges by means of simply deploying more antennas per unit area. The random phase multiple access (RPMA) \cite{rpma} technology used in Ingenu\texttrademark enables higher data rates (up to hundreds of thousands bps) than LoRaWAN\texttrademark or SIGFOX\texttrademark, at the cost of higher current consumption. Ingenu\texttrademark is also capable of performing effective bidirectional transmissions by synchronizing the GW and the STAs, which transmit within predefined slots using random backoffs.
		
		% DASH7
		\item \textbf{D7AP:} the DASH7 Alliance Protocol (D7AP) is a full stack specification by the DASH7 Alliance implementing the complete open systems interconnection (OSI) model, including the presentation and application layers, and uses the Bursty Light Asynchronous Stealth Transitive technology\footnote{\textit{Bursty} stands for abrupt data transfer, \textit{Light} refers to limited and small packet size, \textit{Asynchronous} refers to command-response communication without any periodic synchronization, \textit{Stealth} stands for the ability of nodes to operate in trusted environments, and \textit{Transitive} refers to the nodes' transitional behavior.} (BLAST) \cite{weyn2013survey}. The main singularity of this LPWAN technology is that it enables both star and tree topologies to facilitate the management of large networks. In the latter case, STAs not reaching the GW directly transmit to sub-controllers or other STAs, which forward the messages to the GW. Regarding the MAC layer, STAs in D7AP check the channel periodically for possible pull requests or downlink transmissions, increasing the power consumption but reducing the latency. When ready to transmit (or push), STAs are able to request immediate communication with sub-controllers, which follow different periodic listening cycles. In \cite{weyn2015dash7}, authors present the DASH7 Alliance Protocol 1.0. and discuss the implementation of every OSI layer.
		
		%HARE
		% TOREVIEW
		\item \textbf{HARE:} this novel protocol stack is presented in \cite{adame2017hare} as a new LPWAN technology flexible enough to adopt uplink multi-hop communications when proving energetically more efficient. The GW in HARE is responsible for controlling the STAs by means of a beaconing schedule. This centralized approach allows STAs to remain asleep (i.e., in low power modes) the majority of their lifetime, waking up only to listen to beacons for receiving specific commands and/or configurations. HARE also includes a smart power management controller enabling STAs to use the minimum power level required for reaching their next hop, which significantly reduces transmitting energy consumption. Authors state that results gathered from a real testbed show energy savings of up to 15\% when using a multi-hop approach while keeping the same network reliability.
		
	\end{itemize}
	
	\subsection{Topologies in LPWANs} \label{subsection:routing_in_lpwans}
\begin{figure*}
		\centering
		\begin{minipage}[b]{.45\textwidth}
			\includegraphics[scale=0.45]{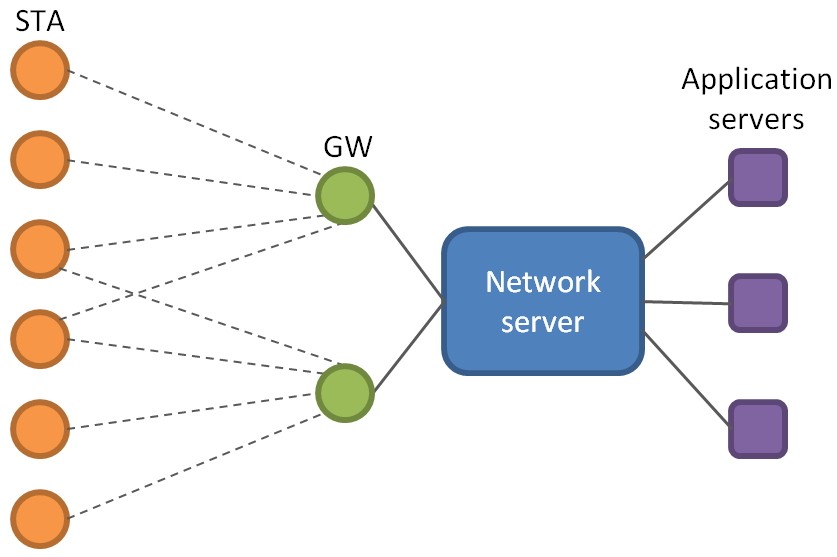}
			\caption{LoRaWAN\texttrademark network topologies are typically laid out in star or star-of-stars.}
			\label{fig:lora_topology}
		\end{minipage}\qquad
		\begin{minipage}[b]{.45\textwidth}
			\includegraphics[scale=0.35]{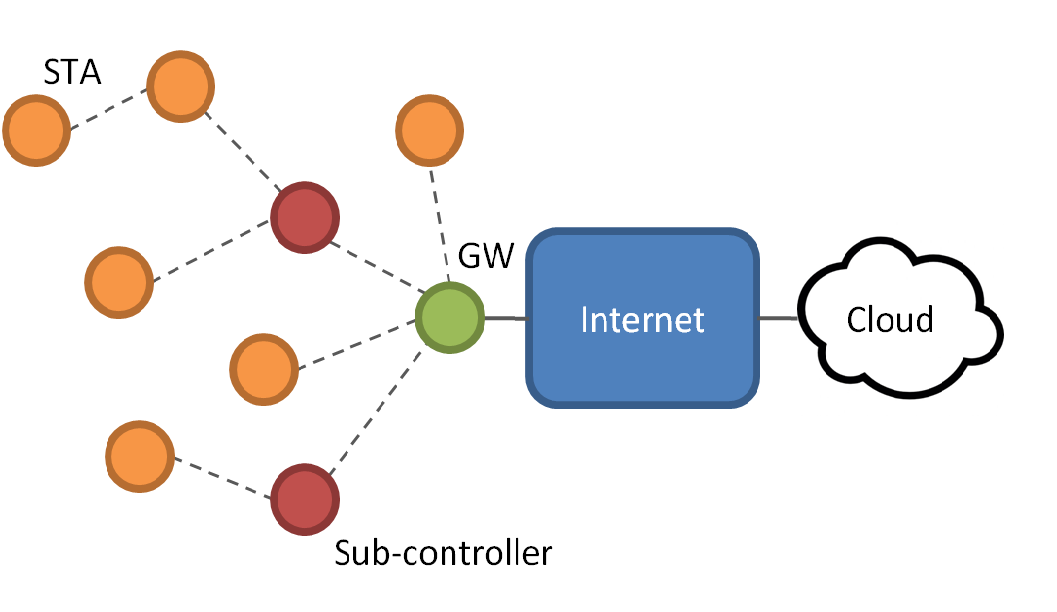}
			\caption{DASH7 networks can be built both in star and tree topologies.}
			\label{fig:d7ap_topology}
		\end{minipage}
	\end{figure*}

	The star topology is predominant in most of LPWAN solutions like LoRaWAN\texttrademark \cite{augustin2016study} (see Figure \ref{fig:lora_topology}) or SIGFOX\texttrademark \cite{sigfoxWhite}, making the STAs to communicate directly with the GW in a single-hop manner. Even though such topology has clear benefits like protocol stack simplification, centralized control, or even infrastructure re-use of traditional cellular networks, it may not be efficient in terms of energy saving. Specifically, in star topology networks, STAs located far from the GW must transmit in high power levels in order to cover the large distance to reach their destination, leading to rapid battery depletion, even more considering the low data rates used in LPWANs which cause larger packet transmission times. This is a critical issue especially in the uplink because STAs are most of the cases battery-powered and therefore energy constrained, while GWs are often plugged to an unlimited power source and do not have such consumption limitations.
	
	Multi-hop alternatives for the uplink in LPWANs technologies have not been profoundly explored yet in networks operating at sub-1GHz. Although there is a small number of solution supporting tree topologies (or multi-hop) like D7AP \cite{weyn2015dash7} (see Figure \ref{fig:d7ap_topology}) or HARE \cite{adame2017hare}, to the best of our knowledge, studies comparing single-hop and multi-hop approaches in LPWANs are not yet available in the literature.
	
	% TOREVIEW: NEW PARAGRAPH FOR REVIEWER NUMBER 2

	Nonetheless, lots of literature have been published in the past years discussing multiple energy-aware routing protocols for wireless sensor networks (WSNs). A detailed survey of routing protocols for WSNs is given in \cite{pantazis2013energy}. Specifically, one of the most relevant approaches for reducing energy consumption and extending the lifetime of WSNs is clustering \cite{amgoth2015energy, tyagi2013systematic}, where nodes are organized in groups (i.e, clusters) in which a certain node (i.e., the cluster head) is responsible for both receiving the sensed data from the rest of nodes within the cluster, and forwarding such aggregated information to a base station in single-hop or multi-hop communication. Despite the huge differences between WSNs and LPWANs, such as the communication range, both of them share a main constraint, which is the limited and often irreplaceable power source of STAs. Therefore, studies on energy-aware routing protocols for WSNs are a valuable starting point when designing new LPWAN routing mechanisms.
	
	%%%%%%%%%%%%%%%%%%%%%%%%%%%%%%%%%%%%%%%%%%%%%%%
	%%% III. DRESG framework and considerations %%%
	%%%%%%%%%%%%%%%%%%%%%%%%%%%%%%%%%%%%%%%%%%%%%%%
	\section{DRESG framework} \label{sec:considerations}
	
	% Battery lifetime challenge and consumption in each state
	In this section we describe the DRESG framework \cite{barrachina2017dresg}, which has been developed for analyzing the impact of enabling multi-hop connections in the uplink on wireless network's consumption. Results in Section \ref{sec:performance_evaluation} show that, in terms of lifetime extension, enabling children-parent routing connections in LPWANs may be a proper alternative against the widely implemented single-hop or star topology.
	
	%%% Energy consumption modeling
	\subsection{Energy consumption modeling}
	STAs consume different amounts of energy per time unit depending on the states they are, which are commonly determined by two sources of energy consumption: microprocessor's ($e_{p}$) and transceiver's ($e_{t}$). Microprocessor states are low power mode (LPM) and processing (CPU), while transceiver states are sleeping (SL), idle (ID), receiving (RX) and transmitting (TX). For the TX state, it is needed to differentiate among each possible transmission power level ($p\in\{1, ..., p_{\text{min}}\}$)\footnote{We use the typical power level notation found in the transceiver datasheets: level 1 for maximum transmission power and level $p_{\text{min}}\geq 1$ for minimum transmission power level. In Table \ref{table:transceivers_power}, the presented power levels follow this notation.} and corresponding transmission power ($P_{\text{tx}}$) due to their different impact on current consumption ($I$).
	\begin{figure}[]
		\centering
		\includegraphics[scale=0.6]{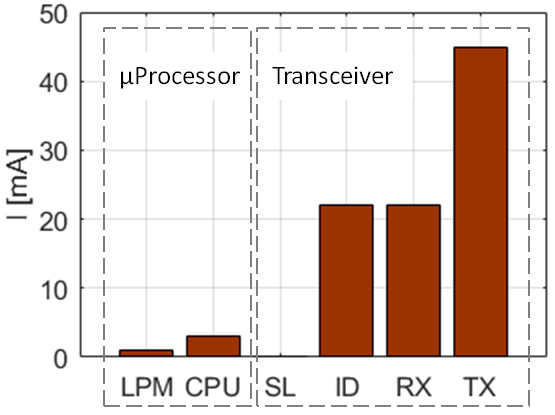}
		\caption[]{Current consumption in an STA composed of a CC2538 ARM Cortex-M3 microprocessor and a CC1200 radio transceiver operating at maximum power level.}
		\label{fig:states_consumption_detailed}
	\end{figure}
	
	The total energy consumed by an STA, $e=e_p+e_t$, can be estimated by measuring the time spent in each of the possible states, i.e.,
	\begin{align}	
	e_p&=\Big(t_{\text{lpm}}I_{\text{lpm}}+t_{\text{cpu}}I_{\text{cpu}}\Big)V_{\text{DD}}\label{eq:e_p}\text{, and}\\	e_t&=\Big(t_{\text{sl}}I_{\text{sl}}+t_{\text{id}}I_{\text{id}}+t_{\text{rx}}I_{\text{rx}}+\sum_{p=1}^{p_{\text{min}}}t_{\text{\text{tx}},p}I_{\text{tx}}(p)\Big)V_{\text{DD}}\label{eq:e_t}\text{,}
	\end{align}
	where $t_{\text{lpm}}$, $t_{\text{cpu}}$, $t_{\text{sl}}$, $t_{\text{id}}$ and $t_{\text{rx}}$ are the time periods the STA is in LPM, CPU, SL, ID and RX states, respectively, and $I_{\text{lpm}}$, $I_{\text{cpu}}$, $I_{\text{sl}}$, $I_{\text{id}}$ and $I_{\text{rx}}$ are the corresponding current consumptions. The time and current consumption in TX state at power level $p$ are defined by $t_{\text{tx},p}$ and $I_{\text{tx}}(p)$, respectively. The nominal voltage is defined by $V_{\text{DD}}$.
	
	As shown in Figure \ref{fig:states_consumption_detailed}, the most current consumption states are TX, RX and ID. The time spent in TX and RX states depends on the packet length, the number of packets to be transmitted, and the data rate. Instead, the time in ID state (i.e., transceiver active but neither transmitting nor receiving) highly depends on the Medium Access Control (MAC) layer. It is common in the literature to consider $I_{\text{id}} \approx I_{\text{rx}}$ for simplicity.
	
	Regarding the transmission power and data rate, we define a \textit{transmission configuration} as the ordered pair $(P_{\text{tx}},s_{\text{tx}})$ corresponding to the uplink child-parent connection. Depending on the node's transceiver, one or more transmission power and data rate levels may be available (or programmable), being the maximum data rate dependent on the sensitivity, as the communication range is determined by the link budget, i.e., the difference between the receiver's sensitivity and the transmission power of the transceiver.
	
	This complex relation among the variables impacting on the transceiver consumption hardens the task of identifying in advance which are the best \textit{transmission configurations} for each of the established routing connections. For example, raising $P_{\text{tx}}$ would also increase $I_{\text{tx}}$, impacting negatively on the energy consumption. However, a higher $s_{\text{tx}}$ can be used when less demanding sensitivities are required, and therefore, $t_{\text{tx}}$ and $t_{\text{rx}}$ could be decreased (e.g., a radio transmitting at 100 kbps remains in TX state approximately twice the time a radio transmitting at 200 kbps), having a positive impact on the energy consumption. Also, the higher the data rate, the less the channel is occupied. Moreover, the transmission power level (and power output) is not usually linear with the corresponding power consumption (see Table \ref{table:transceivers_power}), which hardens even more identifying the most suitable \textit{transmission configuration} beforehand \cite{cc1100,cc1200,Si4464,SX1272}.
	
	\begin{table}[h!]
		\centering
		\begin{tabular}{|c|c|c|}
			\hline
			\textbf{Concept}                                                            & \textbf{Variable} & \textbf{Description}           \\ \hline
			& $D$                          & Distance between the GW and the furthest ring \\ \cline{2-3} 
			& $R$                          & Number of rings                           \\ \cline{2-3} 
			& $c$                          & Tree children ratio                       \\ \cline{2-3} 
			& $B$                          & Number of branches                        \\ \cline{2-3}
			& $b$                          & Branch load		                       \\ \cline{2-3} 
			& $N$                          & Number of STAs                            \\ \cline{2-3} 
			\multirow{-6}{*}{\begin{tabular}[c]{@{}c@{}}DRESG \\ network\\ structure\end{tabular}} & $d$                          & Distance to the GW                        \\ \hline
			& $\vec{\delta}$               & Ring hop lengths                          \\ \cline{2-3} 
			& $\varDelta$                  & Set of ring hop lengths combinations      \\ \cline{2-3} 
			\multirow{-3}{*}{Topology}                                                             & $\Lambda$                    & Connectivity matrix                       \\ \hline
			& $t_{\text{S}}$               & Time in state $S$                         \\ \cline{2-3} 
			& $I_{\text{S}}$               & Current consumption of state $S$          \\ \cline{2-3} 
			& $V_{\text{DD}}$              & Nominal voltage                           \\ \cline{2-3} 
			& $e_{\text{tx}}$              & TX energy consumed by an STA              \\ \cline{2-3} 
			& $e_{\text{rx}}$              & RX energy consumed by an STA              \\ \cline{2-3} 
			& $e$                          & Total energy consumed by an STA           \\ \cline{2-3} 
			& $e_{\text{bt}}$              & Total energy consumed by bottleneck STAs  \\ \cline{2-3} 
			\multirow{-8}{*}{\begin{tabular}[c]{@{}c@{}}Energy \\ consumption\end{tabular}}        & $e_N$                        & Total energy consumed by the network    \\ \hline
			& $P_{\text{tx}}$              & Transmission power                        \\ \cline{2-3} 
			& $p$                          & Transmission power level                  \\ \cline{2-3} 
			& $s_{\text{tx}}$              & Data rate                                 \\ \cline{2-3} 
			& $s$                          & Data rate level                           \\ \cline{2-3} 
			& $C$                          & Set of transmission configurations        \\ \cline{2-3} 
			& $S$                          & Sensitivity                               \\ \cline{2-3} 
			\multirow{-7}{*}{\begin{tabular}[c]{@{}c@{}}Transmission\\ configuration\end{tabular}} & $\text{PL}$                         & Path loss                                 \\ \hline
		\end{tabular}
		\caption{Summary of the main variables used in the DRESG framework.}
		\label{table:variables}
	\end{table}
	
	%%% DRESG description
	\subsection{DRESG description} \label{framework_description}
	
	The DRESG framework is used to evaluate the routing impact on the lifetime\footnote{We consider the lifetime as the time the full network remains operative, i.e., until any STA runs out of battery.} in tree-based deployed networks with STAs spread in distance rings\footnote{STAs spreading in distance rings is a representative case of LPWANs and WSNs. Some applications like agriculture monitoring, intelligent buildings, biodiversity mapping, etc. may be built on networks composed of STAs spread in distance rings around the GW.}. DRESG only considers two sources of transceiver's current consumption (TX and RX) as the TDMA-based MAC presented in \ref{subsection:considerations} is assumed. Every STA in DRESG generates its payload and sends it to its parent, which aggregates its own and all the payloads received from its direct children. Thus, the number of packets to be sent by an STA depends on the amount of payloads received. Similarly, the time a parent STA is in RX state depends on the number of children and the amount of packets they transmit. A parent node could be the GW or another STA, depending on the routing connections established. In Table \ref{table:variables}, a summary of the main variables used in the DRESG framework is shown.
	
	\subsubsection{Network structure}
\begin{figure}[h!]
		\centering
		\includegraphics[scale=0.4]{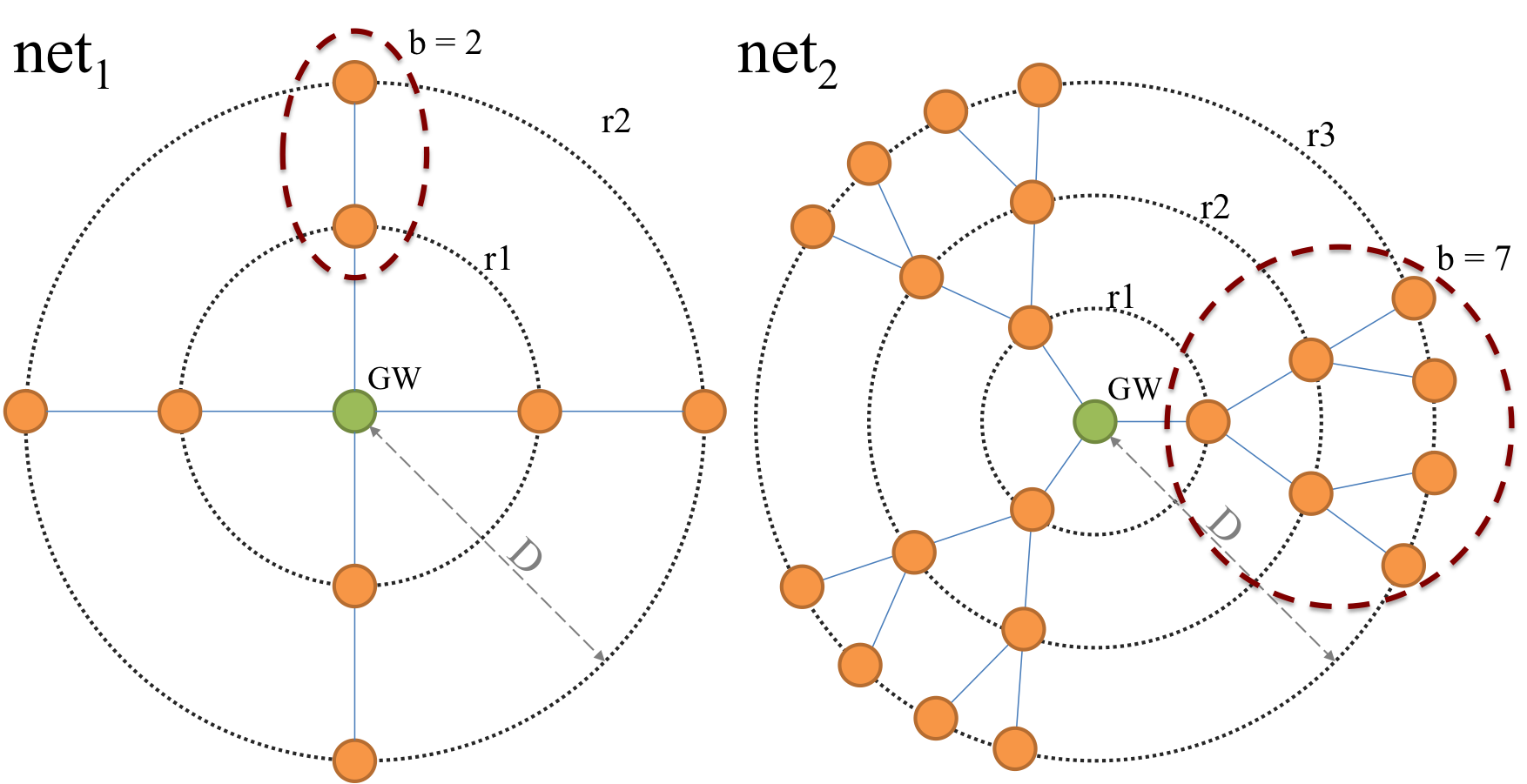}
		\caption{DRESG network structures examples. $\text{net}_1$ = \{$D$, $R = 2$, $c = 1$, $B = 4$\} and $\text{net}_2$ = \{$D$, $R = 3$, $c = 2$, $B = 3$\}. The \textit{branch loads} of networks $\text{net}_A$ and $\text{net}_B$ are 2 and 7, respectively.}
		\label{fig:net_topo}
	\end{figure}

\begin{figure}[h]
		\centering
		\includegraphics[scale=0.45]{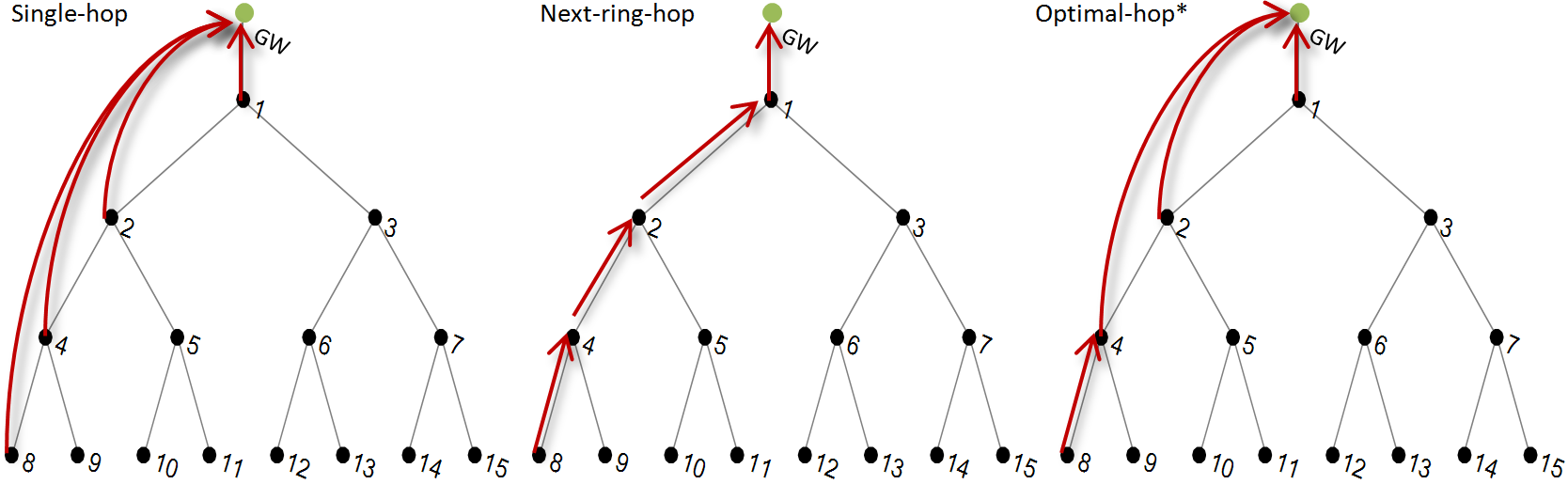}
		\caption{Routing models in a DRESG network structure with $R=4$ and $c=2$. Ring hops (represented as red arrows) in single-hop and next-ring-hop are fixed for any network structure and, in this case, they can be expressed as $\vec{\delta}_{\text{SH}}=(4, 3, 2, 1)$ and $\vec{\delta}_{\text{NRH}}=(1, 1, 1, 1)$, respectively. The optimal-hop connections may vary depending on the network structure and configuration parameters. In this case, the optimal ring hops combination represented in the figure is $\vec{\delta}_{\text{OH}}=(1, 3, 2, 1)$.}
		\label{fig:routings}
	\end{figure}

	In DRESG, STAs are spread in distance rings composing a tree-based network structure that can be defined by the following 4 parameters:
	\begin{itemize}
		\item \textbf{Maximum distance ($\boldsymbol{D}$):} STAs at the furthest ring (i.e., last ring) are placed at distance $D$, which is given by the coverage range provided the GW's transceiver at maximum $P_{\text{tx}}$ and minimum $s_{\text{tx}}$.
		
		\item\textbf{Number of rings ($\boldsymbol{R}$) and distance spreading:} the number of rings in a DRESG network structure is defined by $R$ and STAs belonging to the same ring are located exactly at the same distance to the GW, which is set depending on the selected distance spreading model: \textit{Equidistant}, \textit{Fibonacci}, or \textit{Reverse Fibonacci} (\textit{R-Fibonacci}). \textit{Equidistant} spreading sets the distance of any ring $r$ proportionally to the number of rings, i.e., $d_{\text{equi}}(r)=r(D/R)$. Instead,  \textit{Fibonacci} spreading sets short distances between rings closer to the GW and larger distances for further rings, while \textit{R-Fibonacci} does the opposite. Specifically,
		\begin{align} 
		d_{\text{fibo}}(r)&=\frac{F_{r+1}D}{F_{R+1}} \label{eq:fibo} \text{, and} \\	
		d_{\text{r-fibo}}(r) &= 
		\begin{cases} 
		d_{\text{fibo}}(R-r+1)-d_{\text{fibo}}(R-r) & \textrm{if } r = 1\\
		D  & \textrm{if } r = R\\
		d_{\text{fibo}}(r-1) + d_{\text{fibo}}(R-r+1)-d_{\text{fibo}}(R-r) & \text{otherwise}
		\end{cases} \label{eq:r-fibo},
		\end{align}
		respectively, where $F_n$ is the $n$th number of the Fibonacci sequence. In Figure \ref{fig:distance_rings}, the three aforementioned distance spreading models are plotted for a DRESG network with 10 rings.
		
		\begin{figure}[h!]
			\centering
			\includegraphics[scale=0.6]{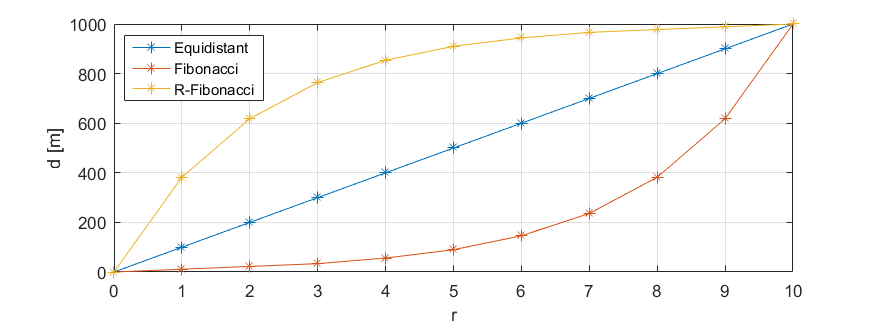}
			\caption{Distance to the GW ($d$) of STAs located in each ring ($r$) of a DRESG network with $R = 10$ and $D=1000$ m.}
			\label{fig:distance_rings}
		\end{figure}
		
		\item \textbf{Tree children ratio ($\boldsymbol{c}$):} number of \textit{tree children}\footnote{We distinguish between \textit{tree children} and \textit{topology children}. On the one hand, \textit{tree children} refers to all STAs of an adjacent higher ring from which an STA (i.e., \textit{tree parent}) may receive packets. On the other hand, \textit{topology children} (children from now on) refer to the STAs in lower adjacent or non-adjacent rings from which an STA actually receives packets. Similarly, \textit{topology parent} (parent from now on) refers to that STA to which a child actually transmits its own packets (after aggregating the ones from its own children) in its way to the GW.} of every STA which does not belong to the last ring. STAs belonging to the last ring have no \textit{tree children}. In Figure \ref{fig:net_topo}, two examples of DRESG network structures are shown. The tree children ratio refers only to the network structure and it is independent to the topology (or routing connections). As shown in Figure \ref{fig:routings}, different topologies may exist for the same DRESG network structure.
		
		\item \textbf{Number of branches ($\boldsymbol{B}$):} a branch is a set of nodes composed of an STA in the \nth{1} ring and its direct and indirect \textit{tree children}. The node load of a branch, or \textit{branch load} ($b$), is defined as the number of STAs in a branch. In DRESG, all branches have the same branch load.
	\end{itemize}

	Hence, the number of STAs ($N$) in an DRESG network can be defined as
	\begin{gather}
	N = B\sum_{r=1}^{R}c^{r-1}\text{,}
	\end{gather}
	being $c^{r-1}$ the number of nodes per branch in ring $r$ for all branches. 
	
	\subsubsection{Configuration parameters}
	DRESG provides several configuration parameters for setting different scenarios.
	\begin{itemize}
		\item \textbf{Data packets and aggregation:} each station in the network generates data payloads of $L_d$ bytes. Each transmitted packet includes a header of $L_h$ bytes. There is a predefined data packet size ($L_{\text{DP}}$), which is the same for any packet transmitted, independently of the number of payloads aggregated\footnote{Some LPWAN solutions such as SIGFOX\texttrademark also predefine a fixed packet size.}. A data packet can contain up to $n^{\max}_p$ data payloads. If the number of payloads aggregated by an station ($n_p$), considering the ones transmitted by its children plus the one generated by the STA itself, is greater than $n^{\max}_p$, then the station will segment the payloads in $n_{\text{DP}}^{\text{tx}}(n_p)$ packets of fixed size $L_{\text{DP}}$, thus including padding bits when the sum of payloads and header is less than $L_{\text{DP}}$, i.e.,
		\begin{align}
		n^{\max}_p &= \left\lfloor{\frac{L_{\text{DP}} - L_h}{L_d}}\right\rfloor \text{, and}\\
		n_{\text{DP}}^{\text{tx}}(n_p) &= \ceil*{\frac{n_p}{n^{\max}_p}}. \label{eq:padding}
		\end{align}
		If $n^{\max}_p = 1$ (i.e., only one payload per data packet is allowed), no children payloads can be aggregated in parent STAs, which will just forward them in different data packets.
		
		\item \textbf{Transceiver model:} depending on the transceiver model selected, a predefined set of power transmission levels are programmable. Also, different transmissions rates may be available depending on the corresponding sensitivity value specified by the transceiver. In general, when transmitting in a high power level, both the power output and current consumption increase. Similarly, for high data rates, sensitivity requirements are expected to increase and large distances may not be reachable.
		
		\item \textbf{Radio propagation model:} the propagation path loss depends on the selected model, which determines the correspondence between reachable distances and required power budgets.
	\end{itemize}
	
	%%% Considerations and assumptions
	\subsection{Considerations and assumptions} \label{subsection:considerations}
	% MAC relevance in energy consumption
	The MAC layer coordinates the transmissions within a shared channel by defining how and when nodes must attempt them. Hence, most of MAC protocols oriented to LPWANs or WSNs save energy by putting the transceiver into SL state as much time as possible, as SL consumes substantially less current than the rest of transceiver's states (see Figure \ref{fig:states_consumption_detailed}). A survey of the different types of low energy operation MAC protocols designed for WSNs is presented in \cite{cano2011low}.
	
\begin{figure}[h]
		\centering
		\includegraphics[scale=0.30]{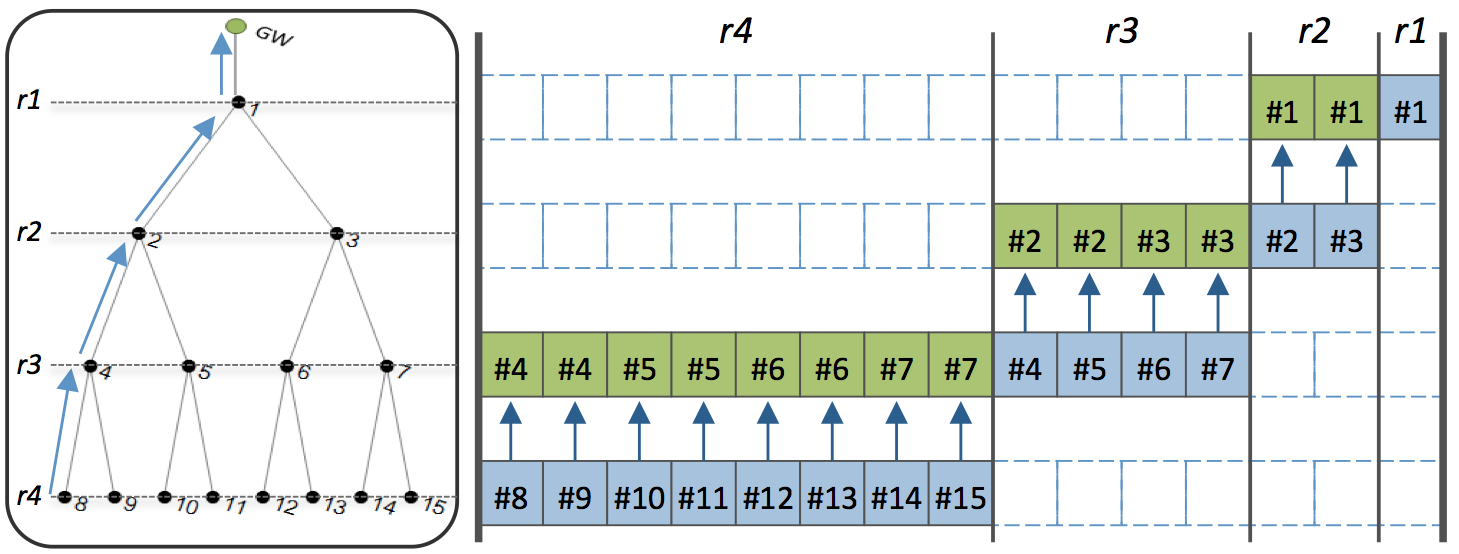}
		\caption{DRESG MAC layer. Example of a network implementing multi-hop communication in the uplink. Blue slots are dedicated to data packet transmissions while green slots represent data packet receptions.}
		\label{fig:mac_4rings}
	\end{figure}

	% Simple and idealistic MAC proposed. 
	There exist several LPWAN technologies, each of them with an specific MAC protocol (e.g., ALOHA in LoRaWAN\texttrademark or RPMA in Ingenu\texttrademark). Therefore, DRESG is focused exclusively on the network topology in order to perform energy consumption analysis regardless of the MAC layer specification. To that aim, we assume a simple and ideal time division multiple access (TDMA) MAC layer with no packet collisions, and where STAs do not need to listen to the channel before transmitting or receiving. Specifically, there are reserved time slots\footnote{The fixed time slots are assigned at the network creation phase.} for packet transmissions for every child-parent routing connection in the network. The size of such slots ($t_{\text{slot}}$) is computed considering the worst case (i.e., lowest data rate and maximum number of aggregated packets) ensuring enough time for every transmission. If a child STA takes less time than $t_{\text{slot}}$ to transmit, the parent and itself are put in SL state right after the packet transmission. That is, parent STAs must listen to the channel exactly the same period of time required by its child to transmit. An example of the MAC layer operation in DRESG is shown in Figure \ref{fig:mac_4rings}.
	
	% Let's focus on the routing
	Apart from the MAC layer, the routing may also have a deep impact on the network lifetime as different power levels and rates can be selected by the transceiver depending on the distance to the parent. In this work, we focus on the routing in the uplink trying to identify ways to minimize the energy consumption at bottleneck nodes (i.e., most energy consuming STAs), while keeping the network with all its STAs fully operative.
	
	The aforementioned assumptions regarding the MAC layer allow us to focus on the analysis of the routing impact on the energy consumption and presume transparent lower layers. Specifically, the DRESG framework is built on the following considerations:
	\begin{enumerate}
		\item \textbf{Negligible idle, sleeping and microprocessor consumption:} using the presented idealized MAC scheme, STAs are not required to check the channel before transmitting or receiving. Instead, when the transceiver is active, it is expected to be in TX or RX states. Therefore, no ID energy consumption is considered in DRESG as $t_{\text{id}}=0$. We also consider $I_{\text{sl}}=0$, as sleeping state current consumption is very low compared to the rest of transceiver states (in the order of nA or \si\micro A depending on several features such as the clock source frequency). Finally, the impact of LPM and CPU states in LPWANs is expected to be very small due to its low current consumption compared to transceiver's states (see Figure \ref{fig:states_consumption_detailed}). In addition, the time spent in such states are expected to be similar in network STAs, as they perform similar processing operations (e.g., gathering data from sensors, buffering, etc.) and they are expected to be most of the time in microprocessor's LPM state.
		\item \textbf{Static routing:} in most of LPWAN envisioned applications, STAs are required to be placed in the same location and not move over time (e.g., agriculture monitoring). Hence, in such scenarios the extra energy cost of implementing a routing protocol (e.g., signaling, update messages, neighbor discovery, etc.) may be very small in comparison of regular data packet transmissions.
	\end{enumerate}
	
	Hence, for an STA in DRESG, the energy consumption obtained from equations (\ref{eq:e_p}) and (\ref{eq:e_t}) is simplified to the sum of the TX and RX energies when transmitting and receiving data packets, respectively, i.e.,
	
	\begin{align}
	e &\simeq e_{\text{tx}} + e_{\text{rx}} \text{, where}\\
	e_{\text{tx}} &\simeq  \sum_{p=1}^{p_{\text{min}}}t_{\text{\text{tx}},p}I_{\text{tx}}(p)V_{\text{DD}} \text{, and}\\
	e_{\text{rx}} &\simeq t_{\text{rx}} I_{\text{rx}}V_{\text{DD}}
	\end{align}
	
\begin{figure}[h]
		\centering
		\includegraphics[scale=0.6]{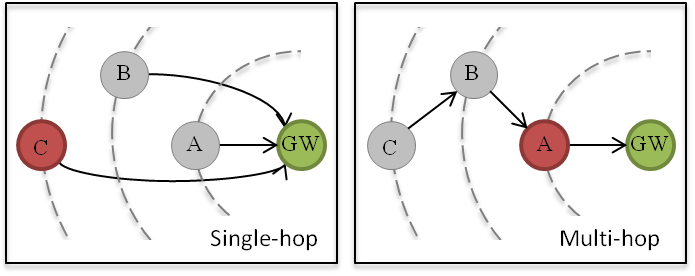}
		\caption{Bottleneck STAs (in red) in single-hop and multi-hop. Depending on the routing model, the bottleneck STAs may vary. While in single-hop topologies bottlenecks are expected to be located far from the GW, in multi-hop ones, STAs with many children may become bottlenecks.}
		\label{fig:routing_example_bottle}
	\end{figure}

	As an example, two simple routing models are shown in Figure \ref{fig:routing_example_bottle}. The bottleneck in the single-hop network is expected to be STA C, located furthest from the GW, due to a higher power level is required in order to reach its destination. One possible way to palliate the consumption of the bottleneck STA is to implement multi-hop in the uplink, letting STAs to reduce their power level when transmitting to closer destinations (C to B and B to A) as shown in the multi-hop network topology. However, generating child-parent routing connections may increase the RX and TX energy on parent STAs as they will have to listen to children packets and forward them. Therefore, a balance in the established connections is needed in order to get a routing model where the energy consumed at bottleneck STAs is minimized.
	
	%%%%%%%%%%%%%%%%%%%%%%%%%%%%%%%%%%%%%%%%%
	%%% IV. ROUTING IMPACT ON CONSUMPTION %%%
	%%%%%%%%%%%%%%%%%%%%%%%%%%%%%%%%%%%%%%%%%
	\section{Routing impact on energy consumption} \label{sec:routing_impact}
	
	%%$ Routing models
	\subsection{Routing models}
	% 'AN LPWAN' INSTEAD 'A LPWAN' http://blog.apastyle.org/apastyle/2012/04/using-a-or-an-with-acronyms-and-abbreviations.html
	The routing model is responsible for determining the path that packets must follow. Thus, it may have high impact on the energy consumed during TX and RX states of STAs composing the network. Hence, it is important to determine which routing connections and \textit{transmission configurations} are optimal in terms of STAs' energy saving. However, this is not a trivial task because of the complex relation among the multiple parameters impacting on consumption. In this subsection, the three routing models implemented in DRESG are presented: single-hop, next-ring-hop, and optimal-hop.	
	
	\subsubsection{Single-hop}
	All STAs (regardless of their ring $r$) transmit their data packets directly to the GW, placed at ring 0. The bottlenecks are expected to be the STAs in the last ring ($r=R$) because they may select higher power levels in order to reach the GW in just one hop. As in single-hop there are no parents in the network, no RX energy is consumed\footnote{The energy consumed by the GW is not considered, as it may be plugged to an unlimited power source or count with energy harvesting mechanisms.}.
	
	\subsubsection{Next-ring-hop}
	
	% TOREVIEW: COMMENT REVIEWER NUMBER 1:	
	Each node in ring $r$ transmits its data packet (or packets) to a parent node placed in the adjacent lower ring, $r-1$. Each parent in the network has exactly $c$ direct children, but the number of indirect children varies depending on the ring it is placed at. In next-ring-hop, the required energy transmission for a single packet is expected to be lower than in single-hop due to distances between source and destination nodes are smaller and lower transmission power levels may be used. However, payload aggregation and listening to children packets can easily generate important bottlenecks at STAs in rings near the GW, specially in ring $r = 1$, due to these STAs aggregate all the payloads of one branch, increasing the RX and TX times in most of the cases, which may lead to rapid battery depletion.

	While single-hop nodes that disconnect from the LPWAN for whatever reason (e.g., internal problem in the node device, battery drain, temporary channel inclemencies, etc.) are often not critical due to the fact that they do not forward any traffic but their own, this is not the case when implementing multi-hop paths. It is important to remark that having nodes forwarding packets from distant rings diminishes the number of node-independent paths, increasing the impact of failures on the overall network performance. Therefore, multi-hop LPWANs must be endowed with routing resilience mechanisms for dealing with such cases. Different routing methods have been designed for maintaining routing paths and facing node failures in WSNs \cite{deng2004robust, attia2007fault}, like periodic reconstruction, multiple routing paths generation, or probabilistic path selection, which could be reused for LPWANs. For example, HARE \cite{adame2017hare} introduces an association mechanism that combines periodic reconstruction and failure detection while trying to maximize the number of node-independent paths through a parent selection weighted metric.
	
	\subsubsection{Optimal-hop}
	Each STA in ring $r$ transmits its data packet (or packets) to a parent in ring $r - \vec{\delta}(r)$, where $\vec{\delta}(r)$ is a vector of $R$ elements representing the hop length of each ring, i.e., the number of rings separating source and destination nodes. In single-hop, $\vec{\delta}(r) = r$ for every ring, as all STAs transmit directly to the GW. Instead, in next-ring-hop, $\vec{\delta}(r) = 1$ for every ring, as all the nodes transmit to a parent in the previous ring. In optimal-hop, $\vec{\delta}(r)$ is set for each ring in a way that the consumption at the bottleneck STAs is minimized. Then, no routing model can perform better than optimal-hop in terms of energy efficiency under DRESG restrictions (e.g., STAs in the same ring must have the same ring destination and same amount of children).
	
	%%% Optimal-hop algorithm
	\subsection{Optimal-hop routing and transmission configurations algorithm}
	The optimal-hop routing and \textit{transmission configuration} algorithm determine two main features given a DRESG scenario: \textit{i}) which are the optimal routing connections, and \textit{ii}) which are the best \textit{transmission configurations} for each of the established routing connections in terms of energy saving. Because the ring hop lengths are not predefined in optimal-hop, the optimal combination of hops may vary depending on the DRESG scenario. The algorithm can be summarized in three main phases depicted below.
	
	\subsubsection{Set of possible ring hops combinations}
\begin{figure}[h]
		\centering
		\includegraphics[scale=0.35]{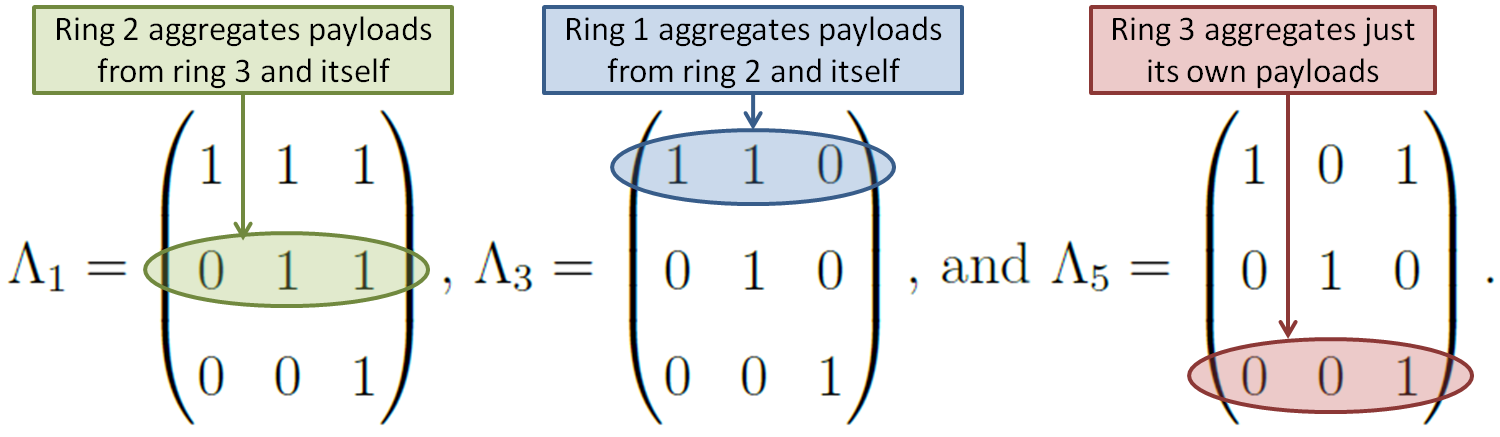}
		\caption{Connectivity matrices corresponding to combinations 1, 3 and 5 from Table \ref{table:hops_combinations}. In combination 3, STAs in the \nth{2} and \nth{3} rings just aggregate their own payloads, while STAs in the \nth{1} ring aggregate the payloads from the \nth{2} ring and their own ones due to $\vec{\delta}(r=3)=3$ (i.e., STAs in the \nth{3} ring transmit directly to the GW), and $\vec{\delta}(r=2)=1$ (i.e., STAs in the \nth{2} ring transmit to the STAs in the \nth{1} ring).}
		\label{fig:connectivity_matrices}
	\end{figure}

	First, we compute the set of ring hops combinations ($\varDelta$), which depends on the number of rings in the network (see algorithm \ref{algorithm_hops}). For example, if $R = 3$, there would be 6 possible combinations as shown in Table \ref{table:hops_combinations}. In order to represent the aggregated payloads in STAs placed at each of the rings corresponding to a certain $\vec{\delta} \in \varDelta$, a connectivity matrix ($\Lambda$) is used. $\Lambda$ is a $R$ x $R$ (0,1)-matrix in which rings and corresponding payload aggregation are represented by rows and columns, respectively. For instance, the connectivity matrices of combinations 1, 3 and 5 from Table \ref{table:hops_combinations} are shown in Figure \ref{fig:connectivity_matrices}.
	
	\begin{algorithm}[h]
		\footnotesize
		\KwData{Number of rings $R$}
		\KwResult{Set of possible ring hops $\varDelta$}
		$s \gets 1$;\\
		\For{$1 \le r \le R!$}
		{\For{$1 \le j \le R!$}
			{$p \gets R!/r!$;\\
				$v \gets \text{mod}(j, r)$;\\
				\If{$v=0$}{
					$v \gets r$;\\
				}
				$\delta_{r}[s: p \cdot j,r] \gets v$;\\
				$s \gets p \cdot j + 1$;\\
			}
			$s = 1$
		}
		\caption{This algorithm computes the set of possible ring hops ($\varDelta$) for a given number of rings ($R$) by considering every possible routing connection among the rings in the network. Any DRESG network has $R!$ hops combinations.}
		\label{algorithm_hops}
	\end{algorithm}
	
	\begin{table}[h]
		\footnotesize
		\centering
		\begin{tabular}{|c|c|c|c|}
			\hline
			\textbf{Combination} & $\boldsymbol{\delta(r=1)}$ &  $\boldsymbol{\delta(r=2)}$ & $\boldsymbol{\delta(r=3)}$ \\ \hline
			\rowcolor[HTML]{CBCEFB}
			1 (next-ring-hop)                   & 1                   & 1                   & 1                   \\ \hline
			2                    & 1                   & 1                   & 2                   \\ \hline
			3                    & 1                   & 1                   & 3                   \\ \hline
			4                    & 1                   & 2                   & 1                   \\ \hline
			5                    & 1                   & 2                   & 2                   \\ \hline
			\rowcolor[HTML]{FFCCC9}
			6 (single-hop)                   & 1                   & 2                   & 3                   \\ \hline
		\end{tabular}
		\caption{Hops combinations for a DRESG network with $R=3$.}
		\label{table:hops_combinations}
	\end{table}
	
	\subsubsection{Bottleneck energy computation}
	Once $\varDelta$ is defined, the algorithm computes the TX and RX energy consumptions of every STA in the network for every combination of hops $\vec{\delta} \in \varDelta$, in all feasible\footnote{A \textit{transmission configuration} is feasible if it ensures reachability between all source and destination pair of nodes in the network. That is, every STA in the network must be able to reach its parent.} set of \textit{transmission configurations} $C=(\vec{P}_{\text{tx}},\vec{s}_{\text{tx}})$, where $\vec{P}_{tx}$ and $\vec{s}_{\text{tx}}$ are vectors of size $R$ whose elements contain the transmission power and data rate of STAs in each ring. On the one hand, every STA in the network has to aggregate the payloads received from its children plus the one generated by itself. Hence, an STA in ring $r$ will aggregate $n_p(r)$ payloads depending on the hops combination and the number of STAs in each ring, i.e.,
	\begin{gather}
	n_p(r) = \sum_{i=r}^{R}\Lambda_{r,i} c^{(i-1)}\text{,}
	\end{gather}
	being $\Lambda_{r,R-i}$ the element placed in row $r$ and column $(R-i)$ of $\Lambda$ matrix. 
	Consequently, the number of data packets to be transmitted by an STA in ring $r$ depends on the number of payloads aggregated and, taking into account the padding effect (see equation (\ref{eq:padding})), it can be expressed as
	\begin{gather} \label{eq:num_pkts_tx}
	n_{\text{DP}}^{\text{tx}}(r) = \ceil*{\frac{ \sum_{i=r}^{R}\Lambda_{r,i} c^{(i-1)}}{n^{\max}_p}}.
	\end{gather}
	Then, the TX energy consumed by an STA in ring $r$ can be straightaway calculated as it is linearly proportional to $n_{DP}(r)$ and to the current consumption ($I_{tx}$) corresponding to the transmission power level ($p$) used during the transmission, i.e.,
	\begin{gather} \label{eq:e_tx}
	e_{\text{tx}}(r) = \bigg(n_{\text{DP}}^{\text{tx}}(r)\frac{L_{\text{DP}}}{s_{\text{tx}}}\bigg) I_{\text{tx}}(p) V_{\text{DD}}\text{.}
	\end{gather}
	
	On the other hand, the total number of data packets an STA in ring $r$ receives, $n_{DP}^{rx}(r)$, depends on the number of packets transmitted by each of its children, which may in turn aggregate payloads from their own children (see equation (\ref{eq:num_pkts_tx})), i.e.,
	\begin{gather}
	n_{\text{DP}}^{rx}(r) = \sum_{i=r}^{R-1}\Lambda_{r,(i+1)}c^{i}\ceil*{\frac{\sum_{j=(i+1)}^{R}\Lambda_{(i+1),j}c^{(j-1)}}{n^{\max}_p}}\text{.}
	\end{gather}
	Then, the RX energy consumed by an STA in ring $r$ can be obtained similarly to the TX energy, but considering in this case the RX current consumption ($I_{rx}$), that is expected to be constant for every packet reception. However, depending on the \textit{transmission configurations} of direct children, different data rates may be used. Thus, 
	\begin{gather}	\label{eq:e_rx}
	e_{\text{rx}}(r) = \bigg(\sum_{i=r}^{R-1}\Lambda_{r,(i+1)}c^{i}\ceil*{\frac{\sum_{j=(i+1)}^{R}\Lambda_{(i+1),j}c^{(j-1)}}{n^{\max}_p}}\frac{L_{\text{DP}}}{s_{\text{tx},R-i}}\bigg)I_{\text{tx}}V_{\text{DD}}\text{,}
	\end{gather}
	where $s_{tx,R-i}$ is the data rate of direct children STAs belonging to ring $R-i$.
	
	Finally, the total energy consumed by an station in ring $r$, $e(r)$, is simply the sum of TX and RX consumptions, i.e., $e(r) = e_{\text{tx}}(r) + e_{\text{rx}}(r)$.
	
	\subsubsection{Optimal ring hops and transmission configurations}
	For every hops combination $\vec{\delta} \in \varDelta$, the algorithm identifies which is the set of \textit{transmission configurations} that generates the less consuming bottleneck STAs ($e_{\text{bt}}$). Afterwards, the tuple composed of the optimal hops combination ($\vec{\delta}^*$) and set of \textit{transmission configurations} ($C^*$) that generates the minimum $e_{\text{bt}}$ is picked, i.e.,
	\begin{gather}
	(\vec{\delta^*}, C^*)=
	\argmin_{%
		\substack{%
			(\vec{\delta}, C) \text{ s.\,t.}\\ 
			C \text{ is feasible}
		}
	}
	e_{\text{bt}}\text{.}
	\end{gather}
	
	%%%%%%%%%%%%%%%%%%%%%%%%%%%%%%%%%%%
	%%%  V. PERFORMANCE EVALUATION  %%%
	%%%%%%%%%%%%%%%%%%%%%%%%%%%%%%%%%%%
	\section{Performance Evaluation}	\label{sec:performance_evaluation}
	
\begin{figure} [h]
		\centering
		\includegraphics[scale=0.5]{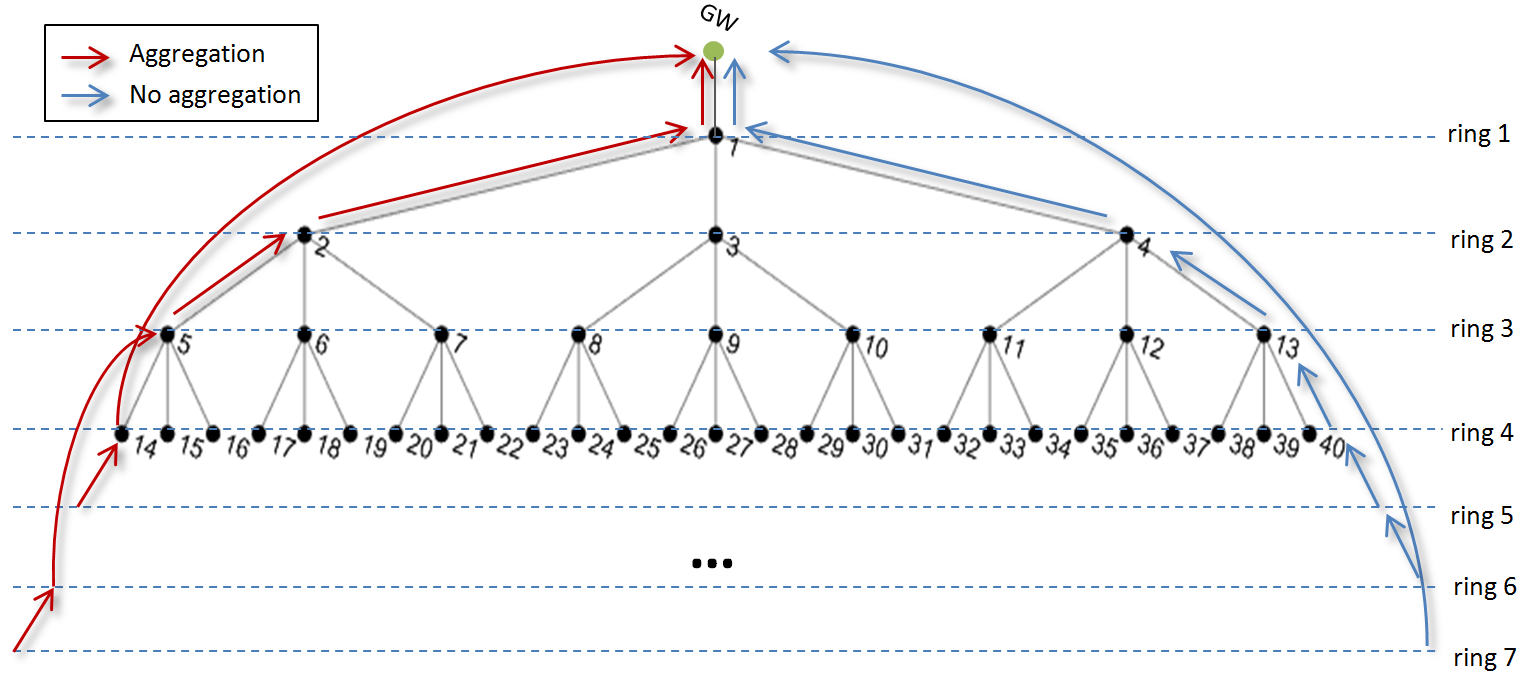}
		\caption{DRESG network structure of Scenario 1.1. The red and blue arrows represent the ring hops obtained through optimal-hop routing when aggregating ($\vec{\delta}_{\text{OH}}$) and when not ($\vec{\delta}_{\text{OH}}^{\text{no agg}}$), respectively. Specifically, $\vec{\delta}_{\text{OH}}=(1, 1, 1, 4, 1, 3, 1)$ and $\vec{\delta}_{\text{OH}}^{\text{no agg}}=(1, 1, 1, 1, 1, 1, 7)$.}
		\label{fig:eval_basic_topo}
	\end{figure}

	% Intro to Scenarios
	In order to evaluate the performance of the optimal-hop routing model in terms of energy saving, several scenarios are considered (see Table \ref{table:eval_scenarios}). From the basic DRESG network structure of Scenario 1.1 (see Figure \ref{fig:eval_basic_topo}), some DRESG network and configuration parameters, i.e., child ratio, number of rings, spreading model, and transceiver model (CC1100 \cite{cc1100}, CC1200\cite{cc1200}, Si4644\cite{Si4464} or SX1272 \cite{SX1272}) are modified to analyze their impact on optimal-hop energy consumption. Tables \ref{table:transceivers_power} and \ref{table:trans_rates_sensitivity} summarize the considered power, data rate, current consumption, and sensitivity specifications of the aforementioned transceivers.
	
	% Fixed scenario parameters
	The rest of parameters are fixed and are the same in every scenario. Specifically, the results presented in this work have been computed considering an 868 MHz carrier frequency and an outdoor path loss model for 802.11ah pico/hot zone deployments defined by \cite{hazmi2012feasibility},
	\begin{gather} \label{eq:pico_model}
	\text{PL}(d) = 23.3+37.6\log_{10}(d)+21\log_{10}\Big(\frac{f}{900 \text{ MHz}}\Big),
	\end{gather}
	where $\text{PL}(d)$ is the path loss in dBs at a distance $d$, and $f$ is the carrier frequency. Besides, all the nodes in the evaluated LPWANs (i.e., both GW and STAs) use the same transceiver model and antennas with transmission gain set to 0 dBi and reception gain set to 3 dBi. The nominal voltage ($V_{\text{DD}}$) is 3 V. Regarding data packets, the parameters implemented in ENTOMATIC EU-project\footnote{ENTOMATIC is an agriculture plague-tracking system that intends to fight the olive fruit fly. It relies on LPWANs where STAs periodically reporting information on pest population density are spread over large olive orchards (1 STA per hectare approximately). Detailed information about the project can be found in the ENTOMATIC main website: \url{https://entomatic.upf.edu/}} are used, where the data payload and header sizes were considered to be $L_{p}=15$ and $L_h=2$ bytes, respectively. The fixed packet size was set to $L_{\text{DP}} = 65$ bytes, allowing to aggregate a maximum number of $n^{\max}_p=4$ payloads per packet. In addition, only one branch per network was considered for simplicity.
	
	\begin{table}
		\footnotesize
		\centering
		\begin{tabular}{|c|c|c|c|c|c|}
			\hline
			\textbf{Scenarios} & $\boldsymbol{R}$ & $\boldsymbol{c}$ & $\boldsymbol{N}$ & \textbf{spreading}   & \textbf{transceiver model}   \\ \hline
			1.1             & 7                & 3                & 1,093            & Equidistant            & CC1200                         \\ \hline
			1.2             & 7                & 2                & 127            & Equidistant & CC1200                         \\ \hline
			2.1 - 2.3             & 7                & 3                & 1,093            & Equid., Fibo., R-Fibo. & CC1200                         \\ \hline
			3.1 - 3.10             & 1 - 10           & 3                & 1 - 29,524       & Equidistant            & CC1100, CC1200, Si4464, SX1272 \\ \hline
			4.1 - 4.10             & 5                & 1 - 10           & 5 - 11,111       & Equidistant            & CC1100, CC1200, Si4464, SX1272 \\ \hline
		\end{tabular}
		\caption{Summary of the network structure and configuration parameters of the DRESG scenarios evaluated.}
		\label{table:eval_scenarios}
	\end{table}
	
	\begin{table}[h]
		\parbox{.45\linewidth}{
			\scriptsize
			\centering
			\begin{tabular}{|c|c|c|c|}
				\hline
				\multicolumn{1}{|l|}{\textbf{Model}} & \multicolumn{1}{l|}{$\boldsymbol{p}$} & \multicolumn{1}{l|}{$\boldsymbol{P_{\textbf{tx}}}$\textbf{[dBm]}} & \multicolumn{1}{l|}{$\boldsymbol{I_{\text{tx}}}$\textbf{[mA]}} \\ \hline
				\multirow{9}{*}{CC1100}           & 1                              & 10.0                                    & 31.1                              \\ \cline{2-4} 
				& 2                              & 7.0                                     & 25.8                              \\ \cline{2-4} 
				& 3                              & 5.0                                     & 20.0                              \\ \cline{2-4} 
				& 4                              & 0.0                                     & 16.9                              \\ \cline{2-4} 
				& 5                              & -5.0                                    & 14.1                              \\ \cline{2-4} 
				& 6                              & -10.0                                   & 14.5                              \\ \cline{2-4} 
				& 7                              & -15.0                                   & 13.0                              \\ \cline{2-4} 
				& 8                              & -20.0                                   & 12.4                              \\ \cline{2-4} 
				& 9                              & -30.0                                   & 11.9                              \\ \hline
				\multirow{16}{*}{CC1200}          & 1                              & 14.0                                    & 45.0                              \\ \cline{2-4} 
				& 2                              & 12.0                                    & 42.0                              \\ \cline{2-4} 
				& 3                              & 10.0                                    & 34.0                              \\ \cline{2-4} 
				& 4                              & 9..0                                    & 33.5                              \\ \cline{2-4} 
				& 5                              & 7.5                                     & 31.0                              \\ \cline{2-4} 
				& 6                              & 5.0                                     & 29.0                              \\ \cline{2-4} 
				& 7                              & 4.0                                     & 27.0                              \\ \cline{2-4} 
				& 8                              & 2.0                                     & 26.0                              \\ \cline{2-4} 
				& 9                              & 0.0                                     & 25.0                              \\ \cline{2-4} 
				& 10                             & -1.5                                    & 24.0                              \\ \cline{2-4} 
				& 11                             & -3.0                                    & 23.0                              \\ \cline{2-4} 
				& 12                             & -5                                      & 22.5                              \\ \cline{2-4} 
				& 13                             & -6.5                                    & 22.0                                \\ \cline{2-4} 
				& 14                             & -8.0                                    & 21.7                              \\ \cline{2-4} 
				& 15                             & -10.0                                   & 21.5                              \\ \cline{2-4} 
				& 16                             & -11.5                                   & 21.0                                \\ \hline
				\multirow{5}{*}{Si4644}           & 1                              & 20.0                                    & 85.0                            \\ \cline{2-4} 
				& 2                              & 16.0                                    & 43.0                              \\ \cline{2-4} 
				& 3                              & 14.0                                    & 37.0                              \\ \cline{2-4} 
				& 4                              & 13.0                                    & 29.0                             \\ \cline{2-4} 
				& 5                              & 10.0                                    & 18.0                             \\ \hline
				\multirow{4}{*}{SX1272}           & 1                              & 20.0                                    & 125.0                             \\ \cline{2-4} 
				& 2                              & 17.0                                    & 90.0                              \\ \cline{2-4} 
				& 3                              & 13.0                                    & 28.0                              \\ \cline{2-4} 
				& 4                              & 7.0                                     & 18.0                              \\ \hline
			\end{tabular}
			\caption{Transceivers' output power and current consumption (based on \cite{cc1100,cc1200,Si4464,SX1272}).}
			\label{table:transceivers_power}
		}
		\hfill
		\parbox{.45\linewidth}{
			\centering
			\scriptsize
			\centering
			\begin{tabular}{|c|c|c|c|c|}
				\hline
				\textbf{Model}	& $\boldsymbol{s_{\textbf{lvl}}}$ & $\boldsymbol{s_{\textbf{tx}}}$\textbf{[kbps]} & $\boldsymbol{S}$\textbf{[dBm]} & $\boldsymbol{I_{\text{rx}}}$\textbf{[mA]}     \\ \hline
				\multirow{3}{*}{CC1100} & 1	&	500                 & -88          	& \multirow{3}{*}{14.4} \\ \cline{2-4}
				& 2	&	250                 & -93           &                       \\ \cline{2-4}
				& 3	&	38.4                & -103          &                       \\ \cline{2-4}
				& 4	&	1.2                 & -110          &                       \\ \hline
				
				\multirow{5}{*}{CC1200} & 1	&	1000                & -97          	& \multirow{5}{*}{19.0}   \\ \cline{2-4}
				& 2	&	500                 & -97          	&                       \\ \cline{2-4}
				& 3	&	100                 & -107          &                       \\ \cline{2-4}
				& 4	&	50                  & -109          &                       \\ \cline{2-4}
				& 5	&	38.4                & -110          &                       \\ \cline{2-4}
				& 6	&	4.8                 & -113          &                       \\ \cline{2-4}
				& 7	&	1.2                	& -122          &                       \\ \hline
				
				\multirow{6}{*}{Si4644} & 1	&	1000                & -88        	& \multirow{6}{*}{10.7} \\ \cline{2-4}
				& 2	&	500                 & -97          	&                       \\ \cline{2-4}
				& 3	&	125                 & -105          &                       \\ \cline{2-4}
				& 4	&	100                 & -106          &                       \\ \cline{2-4}
				& 5	&	40                  & -110          &                       \\ \cline{2-4}
				& 6	&	0.5                 & -126          &          				\\ \hline
				\multirow{8}{*}{SX1272} 	& 1	&	250                	& -97        	& \multirow{8}{*}{10.5} \\ \cline{2-4}
				& 2	&	38.4                & -110          &                       \\ \cline{2-4}
				& 3	&	3.750               & -116        	& 						\\ \cline{2-4}
				& 4	&	18.75               & -119          &                       \\ \cline{2-4}
				& 5	&	9.380               & -122          &                       \\ \cline{2-4}
				& 6	&	1.172               & -131          &                       \\ \cline{2-4}
				& 7	&	0.586               & -134          &                       \\ \cline{2-4}
				& 8	&	0.293               & -137          &          				\\ \hline
			\end{tabular}
			\caption{Transceivers' data rate, sensitivity, and RX current consumption (based on \cite{cc1100,cc1200,Si4464,SX1272}).}
			\label{table:trans_rates_sensitivity}
		}
	\end{table}
	
	%%% Routing models comparison
	\subsection{Routing models comparison}
	
	In Scenario 1.1 we compare the bottleneck consumption for the different routing models presented. Also, we evaluate the energy consumed by the whole network in Scenarios 1.1 and 1.2. In addition, we analyze the effectiveness of implementing payload aggregation in optimal-hop compared to implementing only packet forwarding (i.e., no aggregation) in such scenarios. 
	
	%% Energy consumption per node
	\subsubsection{Ring's energy consumption}
	
\begin{figure}
		\begin{subfigure}{\linewidth}
			\centering
			\includegraphics[scale=0.52]{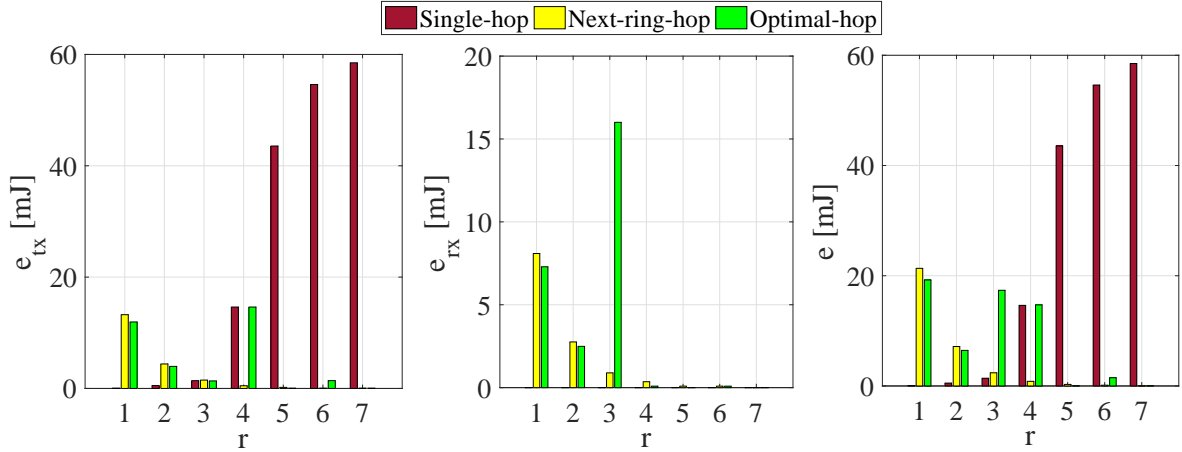}
			\caption{TX, RX and total energy consumed in Scenario 1.1. The total energy ($e$) consumed by single-hop's bottleneck STAs (placed in the \nth{7} ring) is the highest one, while optimal-hop's bottleneck (at the \nth{1} ring) is the less consuming one.}
			\label{fig:exp1_node}
		\end{subfigure}
		\\
		\begin{subfigure}{\linewidth}
			\centering
			\includegraphics[scale=0.54]{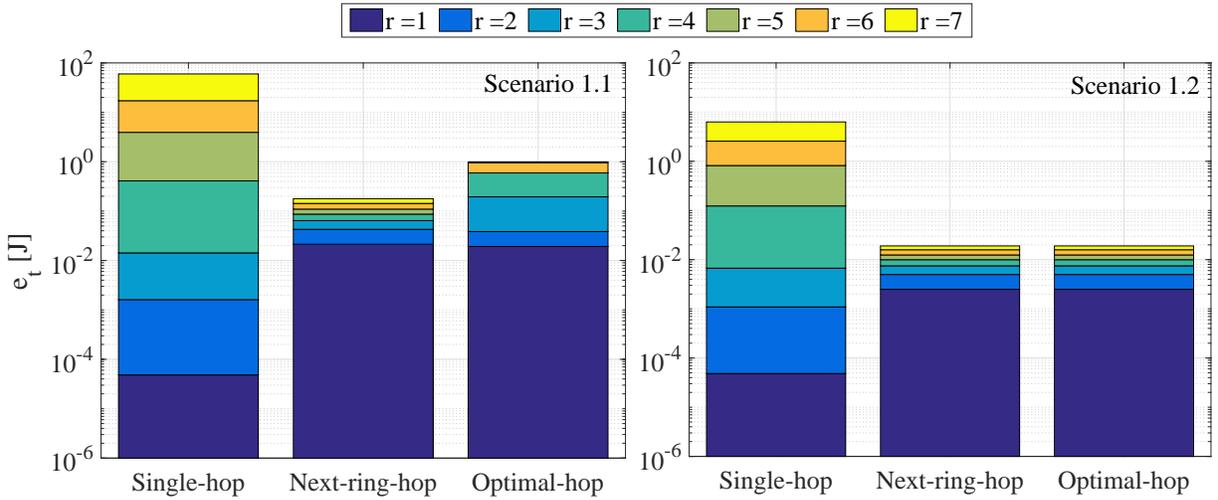}
			\caption{Total energy consumed by the networks in Scenarios 1.1 and 1.2. Through node load balancing among rings, optimal-hop achieves an optimal consumption reduction at bottleneck STAs.}
			\label{fig:exp1_total}
		\end{subfigure}
		\caption{Ring's energy consumption results.}  
	\end{figure}

	As shown in Figure \ref{fig:exp1_node}, for the given DRESG network structure in Scenario 1.1 (with $c=3$ and $R=7$), single-hop's bottleneck is located at the \nth{7} ring, while the energy consumption in optimal-hop's bottleneck (placed at the \nth{1} ring in this case) is clearly minimized, being most appropriate routing model for extending the network lifetime. As expected, next-ring-hop's bottleneck is located also in the \nth{1} ring due to it aggregates and listens to all the payloads of the branch.
	
	In Figure \ref{fig:eval_basic_topo}, we show that optimal-hop balances the traffic by combining next-ring hops from the \nth{3} ring to the GW and establishing routing connections of different hop length for the rest of STAs (e.g., single-hop in the \nth{4} ring). This way, the bottleneck consumption in the \nth{1} ring when using optimal-hop is smaller than when using next-ring-hop, due to the fact that the STA in such ring does not spend so much time in TX and RX states aggregating and listening to packets from STAs in the \nth{2} ring. Nonetheless, the drawback of connecting the \nth{4} ring and the GW directly is the extra energy consumed by STAs in that ring, which is much larger in optimal-hop than in next-ring-hop because a higher transmission power level and a lower data rate is used in order to cover the distance to the GW. In addition, in optimal-hop, STAs in the \nth{3} ring are the most RX energy consuming as they have to listen to packets from STAs in the \nth{6} ring.
	
	In Figure \ref{fig:exp1_total}, the total energy consumed by all the STAs ($e_N$) in Scenarios 1.1 ($c=3$) and 1.2 ($c=2$) are plotted for each routing model. As the furthest ring of the network ($r=7$) is the one with most STAs (729 and 64 for $c=3$ and $c=2$, respectively), in single-hop, their consumption has a deep impact on $e_N$. However, there are cases (as in Scenario 1.1) where optimal-hop, even though generating less consuming bottlenecks than next-ring-hop, causes higher network energy consumption. Nonetheless, the network consumption may not be a priority parameter if network lifetime is determined by the bottleneck STAs.
	
	%% Aggregation vs. no aggregation
	\subsubsection{Payload aggregation vs. no aggregation}
	
	In order to evaluate the expected effectiveness of implementing payload aggregation, we compare the results of the optimal-hop routing model in Scenarios 1.1 and 1.2 when aggregating (i.e., $n_p^{\text{max}}=4$) and when not (i.e., $n_p^{\text{max}}=1$), making the STAs send as many packets as payloads receive plus their own one. We also consider the single-hop model as baseline.
	
	The first noticeable consequence of not implementing payload aggregation in the network with $c=3$ (i.e., Scenario 1.1) is the modification of the optimal-hop routing connections. As seen in Figure \ref{fig:eval_basic_topo}, when not aggregating, the STA in the \nth{1} ring is no longer able to receive so many data packets without consuming a large amount of energy. Instead, due to the fact that a higher amount of packets are sent in the network when payload aggregation is not enabled, a single-hop connection for STAs in the \nth{7} ring is established, avoiding overloading parent STAs with too many packets receptions and transmissions.
	
	The resulting optimal \textit{transmission configurations} of Scenarios 1.1 and 1.2 are summarized in table \ref{table:exp3_agg} for single-hop and optimal-hop with and without aggregation. We note that in Scenario 1.1 (with $c=3$), in single-hop and optimal-hop without aggregation, STAs in the \nth{7} ring must select the maximum transmission power (level 1/16) and minimum data rate (level 7/7) in order to reach the GW directly ($\delta(7)=7$). In optimal-hop with no aggregation, there is no way of generating less consuming bottlenecks because of the large amount of STAs in such ring, which makes packet forwarding energetically unaffordable. Instead, when aggregating, STAs in the \nth{7} ring are able to transmit to parents located in the \nth{6} ring ($\delta(7)=1$), which are much closer than the GW. However, this load balancing comes at a price of moving the bottleneck to the STA placed at the \nth{1} ring, as it receives a large amount of data packets from children in the \nth{2} ring. On the other hand, in Scenario 1.2 (with $c=2$) the number of STAs in the network is deeply reduced ($N=127$), and optimal-hop is able to establish the same routing connections as next-ring-hop regardless of implementing aggregation or not.
	
	\begin{table}
		\scriptsize
		\centering
		\begin{tabular}{|c|c|
				>{\columncolor[HTML]{FFCCC9}}c |
				>{\columncolor[HTML]{FFCCC9}}c |
				>{\columncolor[HTML]{FFCCC9}}c |
				>{\columncolor[HTML]{FFCCC9}}c |
				>{\columncolor[HTML]{FFCE93}}c |
				>{\columncolor[HTML]{FFCE93}}c |
				>{\columncolor[HTML]{FFCE93}}c |
				>{\columncolor[HTML]{FFCE93}}c |
				>{\columncolor[HTML]{9AFF99}}c |
				>{\columncolor[HTML]{9AFF99}}c |
				>{\columncolor[HTML]{9AFF99}}c |
				>{\columncolor[HTML]{9AFF99}}c |}
			\hline
			\multicolumn{1}{|l|}{\cellcolor[HTML]{EFEFEF}}                                & \cellcolor[HTML]{EFEFEF}                               & \multicolumn{4}{c|}{\cellcolor[HTML]{EFEFEF}\textbf{Single-hop}}                                                                                                                                                            & \multicolumn{4}{c|}{\cellcolor[HTML]{EFEFEF}\textbf{Optimal-hop (no agg.)}}                                                                                                                                                                                & \multicolumn{4}{c|}{\cellcolor[HTML]{EFEFEF}\textbf{Optimal-hop}}                                                                                                                                                                                          \\ \cline{3-14} 
			\multicolumn{1}{|l|}{\multirow{-2}{*}{\cellcolor[HTML]{EFEFEF}\textbf{Scenario}}} & \multirow{-2}{*}{\cellcolor[HTML]{EFEFEF}\textbf{$r$}} & \cellcolor[HTML]{EFEFEF}$\boldsymbol{p}$ & \cellcolor[HTML]{EFEFEF}$\boldsymbol{s_{\text{\textbf{lvl}}}}$ & \cellcolor[HTML]{EFEFEF}$\boldsymbol{\delta}$ & \cellcolor[HTML]{EFEFEF}$\boldsymbol{n_p}$ $ \boldsymbol{(n_{\text{\textbf{DP}}}})$ & \cellcolor[HTML]{EFEFEF}$\boldsymbol{p}$ & \cellcolor[HTML]{EFEFEF}$\boldsymbol{s_{\text{\textbf{lvl}}}}$ & \cellcolor[HTML]{EFEFEF}$\boldsymbol{\delta}$ & \cellcolor[HTML]{EFEFEF}$\boldsymbol{n_p}$ $\boldsymbol{(n_{\text{\textbf{DP}}})}$ & \cellcolor[HTML]{EFEFEF}$\boldsymbol{p}$ & \cellcolor[HTML]{EFEFEF}$\boldsymbol{s_{\text{\textbf{lvl}}}}$ & \cellcolor[HTML]{EFEFEF}$\boldsymbol{\delta}$ & \cellcolor[HTML]{EFEFEF}$\boldsymbol{n_p}$ $\boldsymbol{(n_{\text{\textbf{DP}}})}$ \\ \hline
			& 1                                                      & 5/16                                     & 1/7                                                            & 1                         & 1 (1)                                                                               & 5/16                                     & 1/7                                                            & 1                                             & 364 (364)                                                                          & 5/16                                     & 1/7                                                            & 1                                             & 985 (247)                                                                          \\ \cline{2-14} 
			& 2                                                      & 4/16                                     & 3/7                                                            & 2                         & 1 (1)                                                                               & 5/16                                     & 1/7                                                            & 1                                             & 121 (121)                                                                          & 5/16                                     & 1/7                                                            & 1                                             & 328 (82)                                                                           \\ \cline{2-14} 
			& 3                                                      & 1/16                                     & 4/7                                                            & 3                         & 1 (1)                                                                               & 5/16                                     & 1/7                                                            & 1                                             & 40 (40)                                                                            & 5/16                                     & 1/7                                                            & 1                                             & 109 (28)                                                                           \\ \cline{2-14} 
			& 4                                                      & 1/16                                     & 6/7                                                            & 4                         & 1 (1)                                                                               & 5/16                                     & 1/7                                                            & 1                                             & 13 (13)                                                                            & 1/16                                     & 6/7                                                            & 4                                             & 4 (1)                                                                              \\ \cline{2-14} 
			& 5                                                      & 4/16                                     & 7/7                                                            & 5                         & 1 (1)                                                                               & 5/16                                     & 1/7                                                            & 1                                             & 4 (4)                                                                              & 5/16                                     & 1/7                                                            & 1                                             & 1 (1)                                                                              \\ \cline{2-14} 
			& 6                                                      & 2/16                                     & 7/7                                                            & 6                         & 1 (1)                                                                               & 5/16                                     & 1/7                                                            & 1                                             & 1 (1)                                                                              & 1/16                                     & 4/7                                                            & 3                                             & 4 (1)                                                                              \\ \cline{2-14} 
			\multirow{-7}{*}{1.1}                                                         & 7                                                      & 1/16                                     & 7/7                                                            & 7                         & 1 (1)                                                                               & 1/16                                     & 7/7                                                            & 7                                             & 1 (1)                                                                              & 5/16                                     & 1/7                                                            & 1                                             & 1 (1)                                                                              \\ \hline
			& 1                                                      & 5/16                                     & 1/7                                                            & 1                         & 1 (1)                                                                               & 5/16                                     & 1/7                                                            & 1                                             & 127 (127)                                                                          & 5/16                                     & 1/7                                                            & 1                                             & 127 (32)                                                                           \\ \cline{2-14} 
			& 2                                                      & 4/16                                     & 3/7                                                            & 2                         & 1 (1)                                                                               & 5/16                                     & 1/7                                                            & 1                                             & 63 (63)                                                                            & 5/16                                     & 1/7                                                            & 1                                             & 63 (16)                                                                            \\ \cline{2-14} 
			& 3                                                      & 1/16                                     & 4/7                                                            & 3                         & 1 (1)                                                                               & 5/16                                     & 1/7                                                            & 1                                             & 31 (31)                                                                            & 5/16                                     & 1/7                                                            & 1                                             & 31 (8)                                                                             \\ \cline{2-14} 
			& 4                                                      & 1/16                                     & 6/7                                                            & 4                         & 1 (1)                                                                               & 5/16                                     & 1/7                                                            & 1                                             & 15 (15)                                                                            & 5/16                                     & 1/7                                                            & 1                                             & 15 (4)                                                                             \\ \cline{2-14} 
			& 5                                                      & 4/16                                     & 7/7                                                            & 5                         & 1 (1)                                                                               & 5/16                                     & 1/7                                                            & 1                                             & 7 (7)                                                                              & 5/16                                     & 1/7                                                            & 1                                             & 7 (2)                                                                              \\ \cline{2-14} 
			& 6                                                      & 2/16                                     & 7/7                                                            & 6                         & 1 (1)                                                                               & 5/16                                     & 1/7                                                            & 1                                             & 3 (3)                                                                              & 5/16                                     & 1/7                                                            & 1                                             & 3 (1)                                                                              \\ \cline{2-14} 
			\multirow{-7}{*}{1.2}                                                         & 7                                                      & 1/16                                     & 7/7                                                            & 7                         & 1 (1)                                                                               & 5/16                                     & 1/7                                                            & 1                                             & 1 (1)                                                                              & 5/16                                     & 1/7                                                            & 1                                             & 1 (1)                                                                              \\ \hline
		\end{tabular}
		\caption{Optimal \textit{transmission configurations} for single-hop and optimal-hop with and without aggregation corresponding to the networks of Scenarios 1.1 and 1.2 (with CC1200 transceiver model). This table summarizes the power level ($p$), data rate level ($s_{\text{lvl}}$), hop length ($\delta$), number of payloads ($n_p$) and packets ($n_{\text{DP}}$) to transmit for STAs in each ring ($r$).}
		\label{table:exp3_agg}
	\end{table}
	
\begin{figure}
		\centering
		\includegraphics[scale=0.5]{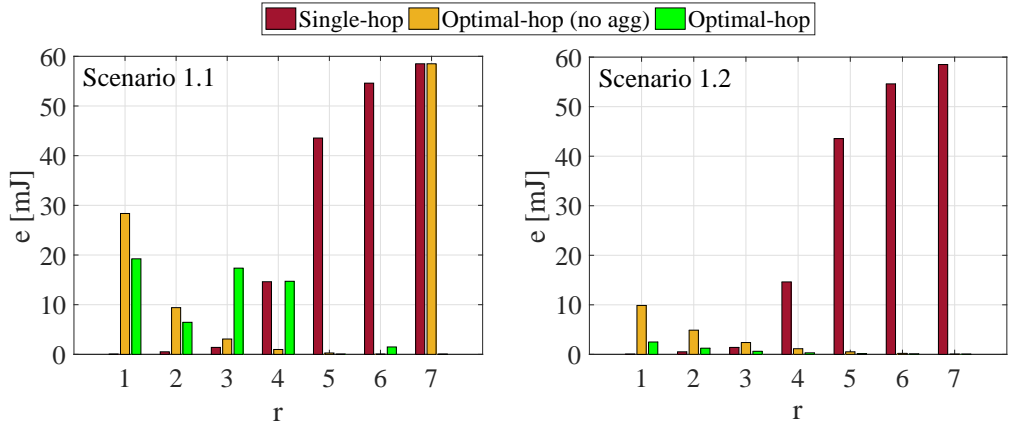}
		\caption{Optimal-hop's payload aggregation vs. no aggregation in networks from Scenarios 1.1 and 1.2. There is a clear improvement in energy saving when aggregating payloads (in green).}
		\label{fig:exp1_aggregation}
	\end{figure}

	In Figure \ref{fig:exp1_aggregation}, the energy consumed by the STAs in each ring is shown. As expected, optimal-hop performs better in terms of lifetime when aggregating (74.7\% and 32.2\% of bottleneck's consumption reduction with respect to optimal-hop without aggregation for $c=2$ and $c=3$, respectively), as more payloads can be included in one single packet. We note that in optimal-hop, the bottleneck ring may not remain the same in optimal-hop when implementing aggregation and when not, as it occurs in Scenario 1.1. Also, even without payload aggregation, optimal-hop outperforms single-hop in terms of bottleneck consumption in Scenario 1.2 (83.1\% of bottleneck's consumption reduction with respect to single-hop). Instead, in Scenario 1.1, such bottleneck consumption cannot be decreased due to the large amount of packets in the network. Nonetheless, optimal-hop without aggregation keeps balancing the traffic in a way that STAs in the \nth{5} and \nth{6} rings consume less than in single-hop, reducing the total energy consumed by the network as a result.
	
	%%% Distance spreading effect
	\subsection{Distance spreading effect}
\begin{figure}
		\centering
		\includegraphics[scale=0.48]{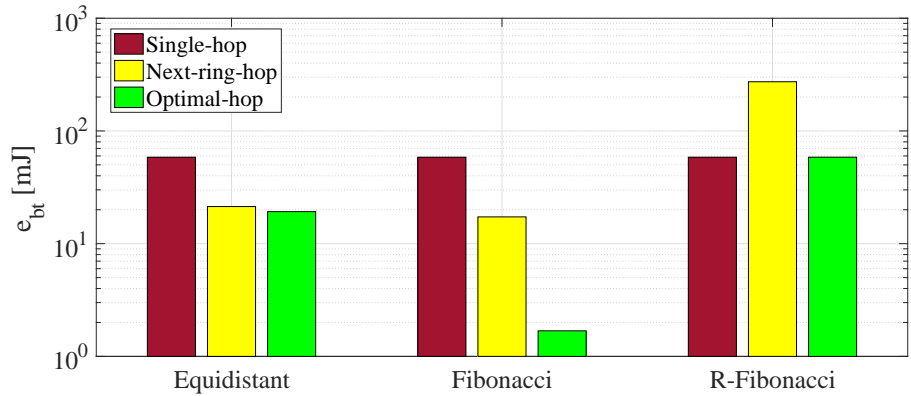}
		\caption{Bottleneck energy in single-hop, next-ring-hop and optimal-hop for each type of ring distance spreading model (Scenarios 2.1 - 2.3). Independently of the distances among rings, single-hop's bottlenecks are always placed in $r=R$.}
		\label{fig:exp2_btle_distances}
	\end{figure}

	Regardless of the distance spreading model selected, single-hop's bottlenecks are always placed at the furthest ring and consume the same amount of energy as they must transmit at maximum transmission power and minimum data rate in order to reach the GW. Also, as shown in Figure \ref{fig:exp2_btle_distances}, there is not a routing model causing larger network lifetime than optimal-hop. In addition, we note the deep effect on the bottleneck's consumption that the distance spreading model may have on next-ring-hop and optimal-hop.
	
	Distances among consecutive rings determine the feasible \textit{transmission configurations}, requiring high transmission powers and lower data rates for large distances, and vice versa for shorter hops. Fibonacci spreading, which sets small distances between rings closer to the GW and larger distances between further rings (see \eqref{eq:fibo}), provides the greatest optimal-hop improvement regarding energy saving, as its bottleneck consumption is the smallest one. Instead, in R-Fibonacci, with large distances among the first rings (see (\ref{eq:r-fibo})), optimal-hop cannot reduce the energy consumption in the last ring and a direct connection to the GW is required.
	
	This occurs due to, in optimal-hop, STAs located in rings close to the GW often aggregate payloads from further rings. Hence, as in R-Fibonacci the first rings are further from the GW, high power levels and low data rates are needed to reach it, increasing the current consumption and transmission time. Instead, in Fibonacci spreading, the STAs aggregating most packets are closer to the GW and are able to select lower transmission power levels and higher data rates accordingly.
	
	%%% Network structure impact
	\subsection{Network structure impact}
\begin{figure}
		\centering
		\includegraphics[scale=0.50]{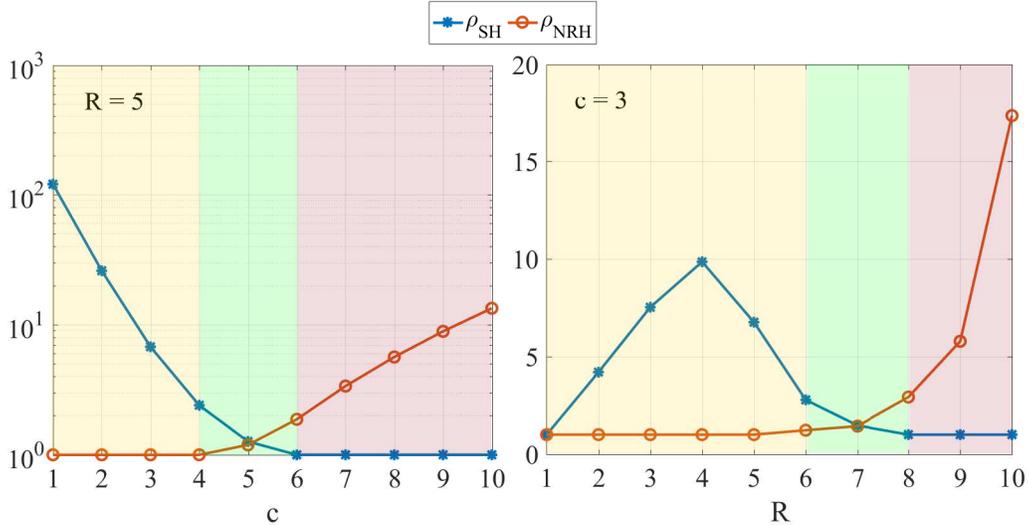}
		\caption{DRESG network structure variation. In crowded networks $\rho_{NRH}$ increases and $\rho_{SH}$ tends to one. The opposite happens for networks with smaller branch loads, where  $\rho_{SH}$ is high and $\rho_{SH}$ approximates to 1. Yellow areas refer to regions where next-ring-hop is optimal, while red areas refer to regions where connections from the last ring to the GW are optimal. In green, other multi-hop topologies generated through the optimal-hop algorithm are the most energy efficient.}
		\label{fig:exp3_topo_vary}
	\end{figure}

	In this subsection, we analyze the impact of the tree children ratio and the number of rings on the optimal-hop performance. To that aim, results shown in Figure \ref{fig:exp3_topo_vary} consider wide ranges of children ratios ($c\in\{1, 2, ..., 10\}$) and number of rings ($R\in\{1, 2, ..., 10\}$) for a SX1272 transceiver with LoRa\textsuperscript{\textregistered} technology. The optimal-hop performance can be measured considering the improvement ratios against single-hop ($\rho_{\text{SH}}$) and next-ring-hop ($\rho_{\text{NRH}}$), i.e., 
	\begin{gather}
	\rho_{\text{SH}} = \frac{e_{\text{bt}}^{\text{SH}}}{e_{\text{bt}}^{\text{OH}}}\text{, and }
	\rho_{\text{NRH}} = \frac{e_{\text{bt}}^{\text{NRH}}}{e_{\text{bt}}^{\text{OH}}}\text{,}
	\end{gather}
	where $e_{\text{bt}}^{\text{SH}}$, $e_{\text{bt}}^{\text{NRH}}$, and $e_{\text{bt}}^{\text{OH}}$ are the energies consumed at the bottleneck STAs in single-hop, next-ring-hop and optimal-hop, respectively. 
	
	As shown in Figure \ref{fig:exp3_topo_vary}, for crowded networks, i.e., with high children ratio ($c$) and/or number of rings ($R$), $\rho_{\text{SH}}$ tends to 1 as optimal-hop establishes a direct connection between STAs in the last ring and the GW, due to aggregating such amount of payloads in parent rings is highly energy consuming. Instead, for networks with smaller number of STAs, $\rho_{\text{SH}}$ increases as payload aggregation in parent STAs allows reducing the consumption in far rings. The peak of $\rho_{\text{SH}}$ curve at $R=4$ is the consequence of \textit{i}) having a larger set of hop combinations with respect $R=3$, and \textit{ii}) reducing $\frac{2}{3}$ the distance among rings (letting STAs use lower transmission powers as parents are close), \textit{iii}) while STAs in the \nth{1} ring still have an small amount of children and payload aggregation is acceptable in terms of energy consumption.
	
	Regarding the improvement ratio against next-ring-hop, the opposite occurs: $\rho_{\text{NRH}}$ increases as networks get more crowded due to the fact that, in next-ring-hop, STAs in the first ring have as many children (direct and indirect) as STAs are in a branch, generating highly consuming bottlenecks. Instead, $\rho_{\text{NRH}}$ approximates to 1 (i.e., optimal-hop sets next-ring-hop routing connections) for networks with lower amount of STAs, as routing in the uplink allows transmitting to closer STAs in lower power levels and higher data rates.
	
	%%% Transceivers analysis
	\subsection{Transceiver analysis}
	
\begin{figure}
		\centering
		\begin{subfigure}{\linewidth}
			\centering
			\includegraphics[scale=0.49]{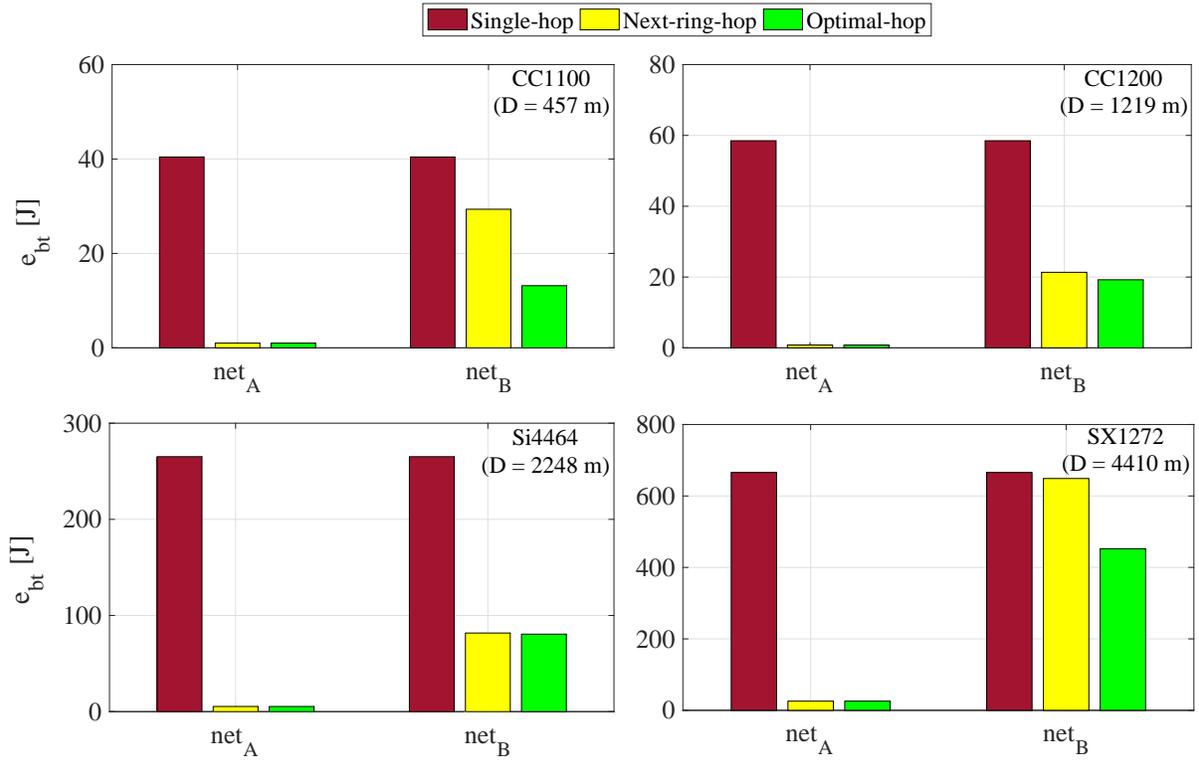}
			\caption{Transceivers' bottleneck consumption in single-hop, next-ring-hop and optimal-hop. Two network structures are evaluated: $\text{net}_A$ = \{$R=5$, $c=2$\} and $\text{net}_B$ = \{$R=7$, $c=3$\} with $N=31$ and $N=1093$ STAs, respectively.}
			\label{fig:exp4_transceivers}
		\end{subfigure}
		\\
		\begin{subfigure}{\linewidth}
			\centering
			\includegraphics[scale=0.55]{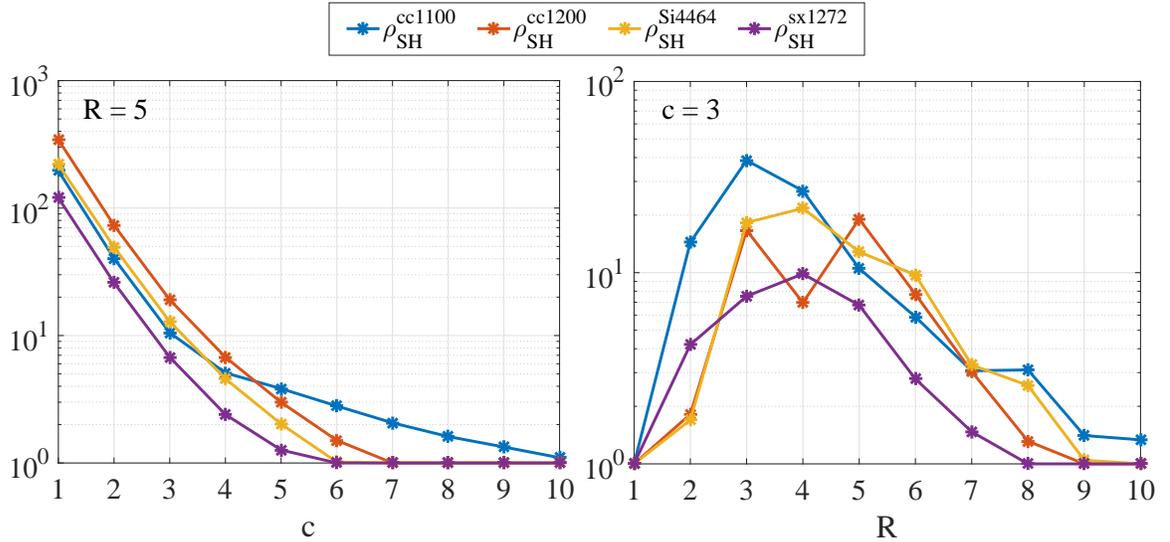}
			\caption{DRESG network structure variations for different transceiver models.}
			\label{fig:exp4_transceivers_and_topo}
		\end{subfigure}
		\caption{Transceivers' consumption results.}  
	\end{figure}

	% Presenting transceivers analysis
	Regarding the transceiver model, available transmission power levels, data rates and sensitivities are critical when it comes to energy saving. In Figure \ref{fig:exp4_transceivers} the bottlenecks' energy consumption in two DRESG network structures with different number of STAs is plotted for four transceiver models: CC1100, CC1200, Si4644 (high-performance, low-current transceiver for SigFox), and SX1272 (low power transceiver featuring LoRa\textsuperscript{\textregistered}).
	
	% Optimal-hops the best for all transceivers. Specially in netA. 
	We note that for all the transceiver models, optimal-hop's bottleneck consumption is the minimum one, being a promising theoretical routing alternative to extend LPWANs' lifetime. Furthermore, special mention should be made of optimal-hop's improvement on network $\text{net}_A$, with $N=31$, where establishing child-parent connections in the uplink highly reduces the bottleneck consumption (more than 96\% of reduction for all the transceivers). However, even having a large number of STAs, in $\text{net}_B$, optimal-hop stills provides a clear improvement in terms of bottleneck energy saving in all the cases. 
	
	% SX1272 suffers the most
	SX1272 is the transceiver providing the largest coverage range (i.e., $D=4410$ meters considering the outdoor path loss model for 802.11ah pico/hot zone deployments) with highest transmitting power (20 dBm and 125 mA current consumption) and lowest sensitivity (-137 dBm at 293 bps). Nonetheless, such range comes at a big price as, in single-hop, extremely consuming bottlenecks are generated in the last ring due to high current consumption and long transmission times.
	
	% Transceivers and deployment variation
	In Figure \ref{fig:exp4_transceivers_and_topo}, we see a similar behavior for different DRESG scenarios regarding the optimal-hop improvement against single-hop. Bottlenecks generated by optimal-hop are in general less consuming (equally consuming in the worst cases, i.e., $\rho_{SH}=1$) than in single-hop, specially for not too crowded networks (i.e., networks with small child ratio or number of rings). The increase of $\rho_{SH}$ from the \nth{1} to the \nth{3}/\nth{4} ring is due to the distance reduction among rings while STAs in first rings still have a small amount of children, and payload aggregation is still energy efficient. Hence, independently of the analyzed transceiver model, the promising optimal-hop performance in terms of energy saving leads to envisage multi-hop connections in the uplink as a powerful way for extending wireless networks' lifetime, specially in LPWANs.
	
	%%%%%%%%%%%%%%%%%%%%%%%%%
	%%%  VI. CONCLUSIONS  %%%
	%%%%%%%%%%%%%%%%%%%%%%%%%
	\section{Conclusions and future work}	\label{sec:conclusions}
	
	% 1 an 2 - LPWANs and single-hop routing
	LPWANs can provide long range communications and low power operation, being a convenient and promising solution for lots of existing and envisioned IoT applications. Nonetheless, most of LPWAN solutions rely on star (or single-hop) topologies, where STAs placed far from the GW may rapidly run out of battery due to required high transmission power levels and low data rates.
	
	% 3, 4 - DRESG framework and Routing impact on energy consumption
	In this work we have presented DRESG, a framework for analyzing the impact on energy consumption of different routing models in several theoretical but representative LPWAN scenarios. The most energetically efficient routing paths were obtained through the optimal-hop and \textit{transmission configurations} algorithm, which identifies the connections among rings that generate the minimum consumption at bottleneck STAs, and what are the most efficient transmission power and data rate levels for every established routing connection. DRESG also implements the single-hop and next-ring-hop routing models, which were used as baselines for comparing the improvement in terms of energy savings with respect to optimal-hop.
	
	% 5 - Performance evaluation
	Results show that optimal-hop topologies lead to higher network lifetimes than with single-hop, thus reducing significantly the consumption of STAs located far from the GW in LPWANs with up to thousands of STAs (more than 96\% of reduction for specific network structures with limited number of STAs). Instead, for networks with a larger number of STAs, single-hop connections from the last ring to the GW are required in order to avoid generating huge bottlenecks at parent STAs. Results also show that such improvements are deeply enhanced with data aggregation (up to 74\% of improvement with respect to the use of no aggregation in the studied cases), and that the benefits of multi-hop communication in the uplink are similar among the considered transceiver models.

	% Encouraging results and next steps
	In conclusion, the obtained results lead to foresee multi-hop communication in the uplink as a promising routing alternative for extending LPWANs' lifetime. The next challenge is to explore a real implementation of a protocol able to converge to optimally (or pseudo-optimally) energy efficient topologies with the minimum consumption overhead possible. Even though such protocol implied extra energy consumption, the presented DRESG results show that in most of the evaluated scenarios there exist huge energy saving budgets to exploit when optimal-hop topologies are used.
	
	\section*{Acknowledgments}
	This work was partially supported by the ENTOMATIC FP7-SME-2013 EC project (605073). It has also been funded by the Catalan government through the project SGR-2014-1173.
	
	\bibliographystyle{unsrt}
	\bibliography{bib}
	
\end{document}